\RequirePackage{lineno}
\documentclass[aps,prd,twocolumn,showpacs,preprintnumbers,superscriptaddress,nofootinbib,longbibliography,floatfix,letterpaper]{revtex4-2}
\pdfoutput=1
\usepackage{placeins}
\usepackage{multirow}
\usepackage[utf8]{inputenc}
\usepackage{color}
\usepackage[caption=false]{subfig}
\usepackage{xspace}
\usepackage{graphicx}
\usepackage[pdftex,bookmarks,hidelinks]{hyperref}

\hypersetup{
  colorlinks=true,
  citecolor=blue,
  linkcolor=blue,
  urlcolor=blue
}

\usepackage{amsmath, amsfonts, amssymb}
\usepackage{caption}
\usepackage{makecell}
\usepackage{comment}

\graphicspath{ {graphics/} }

\begin{document}

\title{Machine Learning Assisted Unfolding for Neutrino Cross-Section Measurements with the OmniFold Technique}
\date{\today}
\author{R.~G.~Huang}
\affiliation{Lawrence Berkeley National Laboratory, Berkeley, CA 94720, USA}
\author{A.~Cudd}
\affiliation{University of Colorado Boulder, Boulder, CO 80309, USA}
\author{M.~Kawaue}
\affiliation{Kyoto University, Department of Physics, Kyoto, Japan}
\author{T.~Kikawa}
\affiliation{Kyoto University, Department of Physics, Kyoto, Japan}
\author{B.~Nachman}
\affiliation{Lawrence Berkeley National Laboratory, Berkeley, CA 94720, USA}
\author{V.~Mikuni}
\affiliation{Lawrence Berkeley National Laboratory, Berkeley, CA 94720, USA}
\author{C.~Wilkinson}
\affiliation{Lawrence Berkeley National Laboratory, Berkeley, CA 94720, USA}

\begin{abstract}
The choice of unfolding method for a cross-section measurement is tightly coupled to the model dependence of the efficiency correction and the overall impact of cross-section modeling uncertainties in the analysis. A key issue is the dimensionality used in unfolding, as the kinematics of all outgoing particles in an event typically affect the reconstruction performance in a neutrino detector. OmniFold is an unfolding method that iteratively reweights a simulated dataset, using machine learning to utilize arbitrarily high-dimensional information, that has previously been applied to proton-proton and proton-electron datasets. This paper demonstrates OmniFold's application to a neutrino cross-section measurement for the first time using a public T2K near detector simulated dataset, comparing its performance with traditional approaches using a mock data study.
\end{abstract}

\maketitle

\section{Introduction}
Unfolding is a general term for removing the impact of a measuring device from a measurement~\cite{Cowan:2002in, Prosper:2011zz}. In the context of cross-section measurements in high-energy physics, this can be thought of as the deconvolution of detector smearing and efficiency effects, necessary to present data in a way which does not require detailed information about the particle detector to interpret. Neutrino cross-section measurements typically use binned unfolding methods and a small number of dimensions to present results. Most current results are single-dimensional, but multi-dimensional results of up to three dimensions have been extracted~\cite{pdg_2024}. A simple, but representative formula used to extract a binned cross section $d\sigma/dx_{i}$ as a function of the true kinematic variable $x$ is~\cite{Mahn:2018mai}:
\begin{equation}
  \frac{d\sigma}{dx_{i}} = \frac{\sum_{j} \widetilde{U}^{-1}_{ij} \left(N_{j} - B_{j}\right)}{\Phi_{\nu}T\Delta x_{i} \epsilon_{i}}
  \label{eq:trad_xsec}
\end{equation}
\noindent where $N_{j}$ ($B_{j}$) is the number of selected (predicted background) events in reconstructed bin $j$, $\Phi_{\nu}$ is the total integrated flux, $T$ is the number of target nuclei, $\Delta x_{i}$ is the width of the true bin $i$ (or area/volume for multi-dimensional extractions), $\epsilon_{i}$ is the selection efficiency, and $\widetilde{U}^{-1}_{ij}$ is the unfolding matrix. The unfolding matrix is a regularized inverse of the matrix $U_{ij}$ that encapsulates the effects of detector smearing. The efficiency and the detector smearing are closely related, and in some formulations the efficiency correction is absorbed into the smearing matrix. A number of unfolding methods have been explored in neutrino physics~\cite{Tang:2017rob, T2K:PhysRevD.98.032003, Gardiner:2024gdy}, between which the most important differences are how the result is regularized and how the potential for bias that regularization introduces is handled~\cite{Mahn:2018mai, Avanzini:2021qlx}. One of the most commonly used unfolding methods in neutrino physics, as in high-energy physics more generally, is D'Agostini unfolding~\cite{D'Agostini:1994zf, DAgostini:2010hil}, also known as Iterative Bayesian Unfolding (IBU)\footnote{Although this is somewhat of a misnomer, as it is a fully frequentist method~\cite{kuusela_master}.}. This method has been independently described multiple times~\cite{Vardi01031985, Multhei:1986ps, kuusela_master} and is known in other fields by other names, such as Lucy-Richardson deconvolution~\cite{Richardson:1972hli, Lucy:1974yx}.

Experimental data from neutrino experiments have a number of unique challenges: they typically suffer from low statistics; detector volumes are large with non-uniform detector efficiency and smearing; they use relatively low-fidelity, high-threshold detectors; and, systematic uncertainties are generally large. These issues couple to the unfolding problem in complex ways, which risks degrading the quality of extracted results. Because the detector smearing and efficiency depend on a large number of variables, expressing a smearing matrix with a small number of dimensions implicitly integrates over all of the other relevant variables. This integration relies not only on the detector model in the simulation, but is convolved with a prediction for how signal events will be distributed in those variables, leading to model-dependence and the potential for bias~\cite{Mahn:2018mai, Avanzini:2021qlx}. Additionally, by expressing the result in a small number of dimensions, the result loses much of the richness of the original data, particularly as the binning must be determined in advance using simulation studies. Experiments often extract results in a number of variables, but generally without correlations between the different variables. This has presented difficulties in the use of these measurements, although methods have been proposed to overcome this problem~\cite{Gardiner:2024gdy}. Additionally, the challenges of unfolding have led to discussions of whether to unfold data at all, both in neutrino physics~\cite{Koch:2019jqr} and more generally~\cite{Cousins:2016ksu}.

New unfolding methodologies are possible using machine learning, many of which naturally generalize to high-dimensional spaces, and a number of approaches have been explored~\cite{Arratia:2021otl,Butter:2022rso2,Huetsch:2024quz}. In this work, we apply the OmniFold unfolding method~\cite{Andreassen:2019cjw,Andreassen:2021zzk} to a publicly available simulated dataset from the T2K experiment and compare its performance and relative advantages with respect to IBU.

This paper is organized as follows.  Section~\ref{sec:OF} briefly introduces machine learning-based unbinned unfolding with OmniFold.  The T2K dataset used for numerical results is described in Sec.~\ref{sec:Dataset}.  Our methods for testing different approaches are documented in Sec.~\ref{sec:TestSetup}. Numerical results are presented in Sec.~\ref{sec:results}, and the paper ends with conclusions and outlook in Sec.~\ref{sec:conclusions}.

\section{Unbinned Unfolding with OmniFold}
\label{sec:OF}

Unfolding methods using neural networks and other machine learning tools process continuous data and are thus naturally unbinned~\cite{Arratia:2021otl,Butter:2022rso2,Huetsch:2024quz}. As these tools can also process high-dimensional feature spaces, unfolding many observables simultaneously is readily accommodated.  Existing approaches are based on likelihood-ratio estimation with machine learning classifiers~\cite{Andreassen:2019cjw,Andreassen:2021zzk,Pan:2024rfh} or are based on direct likelihood-estimation with generative machine learning tools~\cite{Datta:2018mwd,Howard:2021pos,Diefenbacher:2023wec,Butter:2023ira,Bellagente:2019uyp,Alanazi:2020jod,Bellagente:2020piv,Vandegar:2020yvw,Backes:2022vmn,Leigh:2022lpn,Ackerschott:2023nax,Shmakov:2023kjj,Shmakov:2024xxx}.  The last couple of years has seen the first unbinned cross-section studies in $ep$ using H1 data~\cite{H1:2021wkz,H1prelim-22-031,H1:2023fzk,H1:2024mox} and in $pp$ using data from LHCb~\cite{LHCb:2022rky}, ATLAS~\cite{ATLAS:2024xxl,ATLAS:2025qtv}, STAR~\cite{Song:2023sxb,Pani:2024mgy}, and CMS~\cite{Komiske:2022vxg,CMS-PAS-SMP-23-008}.  Since all of the existing experimental results use the OmniFold method, we focus on that technique here.

\begin{figure}
    \includegraphics[width=\linewidth]{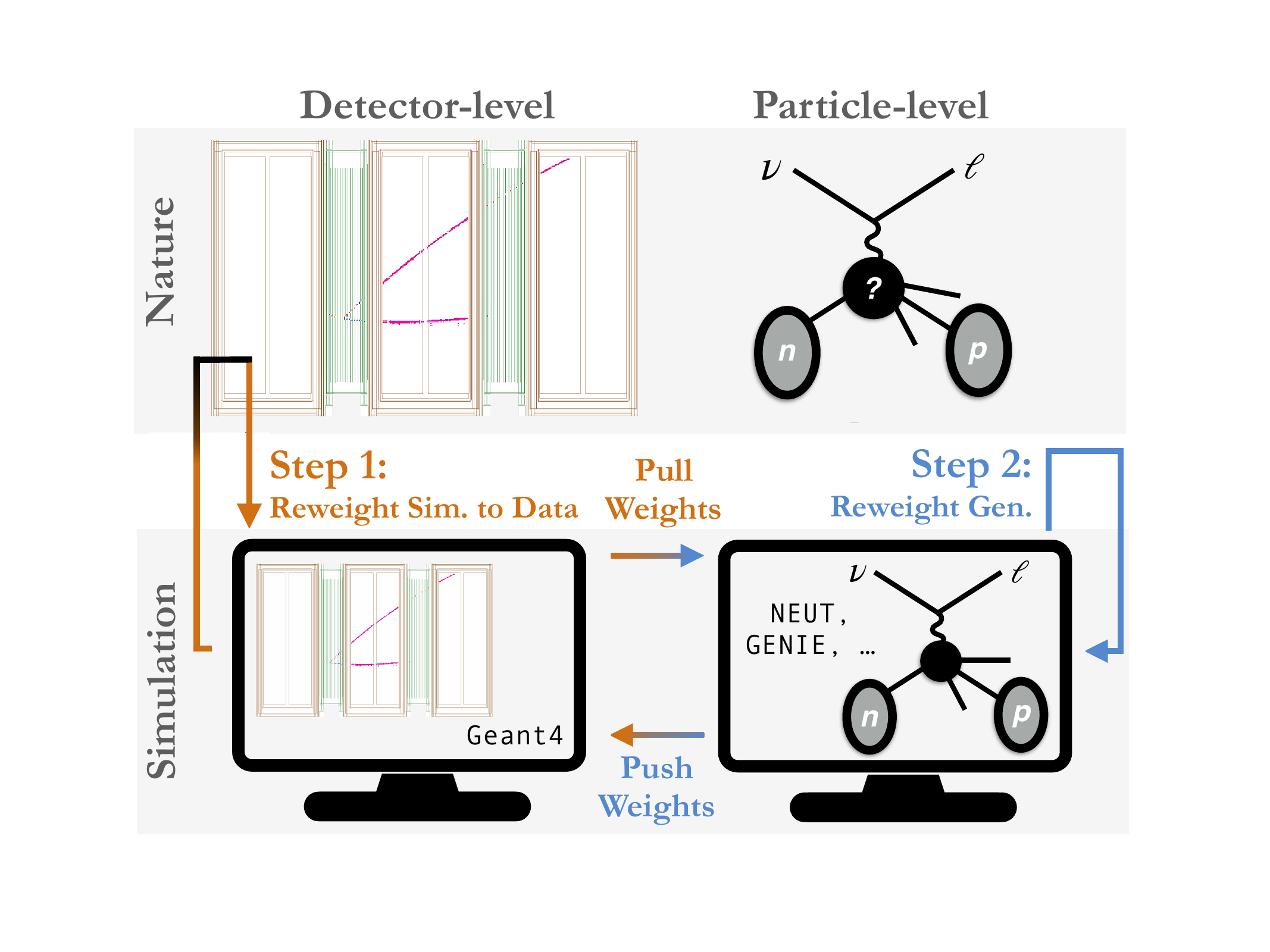}
    \caption{A sketch of the OmniFold method applied to T2K data and simulation, with an example of how a typical charged current neutrino interaction (right) appears in the T2K near detector (left). Step 1 works with detector-reconstructed quantities, comparing simulated detector events against data and reweighting the former to fit the latter. These weights are pulled to step 2, which compares truth-level generator events against themselves with and without the pulled weights to provide reweighting of the true particle-level events.}
    \label{fig:omnifoldschematic}
\end{figure}

OmniFold is an iterative, two-step procedure that is illustrated in Fig.~\ref{fig:omnifoldschematic}. It exploits the manner in which each event from a simulated dataset will have both ``particle-level'' and ``detector-level'' features. Particle-level features refer to the truth-level list of particles that resulted from an interaction along with their associated kinematics, provided by some kind of event generator. Detector-level features refer to the reconstructed quantities one obtains after pushing the particle-level features through the detector simulation and reconstruction algorithms, and these are the quantities one has access to in an actual experiment. In the first step of OmniFold, the detector-level simulated data are reweighted to match the observed data.  Then, these weights are `pulled back' to the particle level by assigning the weight obtained from the detector level in the first step.  Events that do not pass the detector-level event selection are assigned the average weight for a given particle-level phase space region either at the end of the unfolding or at each iteration (comparisons described below).  This pull back induces a new spectrum at particle level, but since the detector response is stochastic, the resulting reweighting is not a proper function of the particle-level phase space.  The second step of OmniFold reweights the starting particle-level simulation to this induced simulation from the pulled back weights.  This weighting function is a proper function of the particle-level phase space by definition.  One can then `push forward' these weights to the detector level and repeat the entire process.  Iterating a finite number of times is a form of regularization, and there are a number of metrics that can be used to decide when to stop.  The final result of the unfolding is a set of particle-level simulated events and a corresponding set of event weights.

While the reweighting can be performed with any method, the algorithm is unbinned when based on machine learning classifiers.  Interpreting classifiers as likelihood ratio estimators,  sometimes called the ``likelihood ratio trick'', is well-known in statistics~\cite{hastie01statisticallearning,sugiyama_suzuki_kanamori_2012} and has been frequently used in particle physics~\cite{cranmer}. 

\section{T2K Public Dataset}
\label{sec:Dataset}

This paper utilizes a public simulated T2K dataset \cite{T2KDataRelease} based on a previous analysis of T2K ND280 near detector data~\cite{T2K:PhysRevD.108.112009}, which was a measurement of charged-current $\nu_\mu$ events that produce zero pions in the final state (referred to as the CC0$\pi$ topology) on a hydrocarbon target. The same dataset or subsets of it have been used in other related T2K publications~\cite{T2K:PhysRevD.98.032003, T2K:2020sbd, T2K:2020jav}. The public simulated dataset represents signal and select background events in both truth and reconstructed space along with associated systematic uncertainties for the flux, detector, and neutrino interaction models. The simulated dataset covers T2K runs 2, 3, 4 and 8 in neutrino mode beam running and contains events corresponding to $17.24\times 10^{21}$ protons on target, roughly a factor of 20 more than the data statistics analyzed in Ref. \cite{T2K:PhysRevD.108.112009}. The ND280 simulation uses the \textsc{neut} package \cite{Hayato:2002sd, Hayato:2009, Hayato:2021heg} to generate neutrino interactions in the detector, \textsc{geant4} \cite{geant4} for particle propagation and energy deposition, and a custom package to simulate the detector electronics response.

ND280 comprises a number of sub-detectors installed inside the refurbished UA1/NOMAD magnet, which provides a 0.2 T field used to measure the charge and momentum of particles. The inner tracker region of ND280 uses three time projection chambers (TPCs) interleaved with two plastic scintillator fine-grained detectors (FGDs), and is surrounded by an electromagnetic calorimeter. The FGDs each contain 1.1 tons of target mass for neutrino interactions and provide tracking of the charged particles coming from the interaction vertex, while the TPCs provide momentum and particle identification. The dataset used here is focused on neutrino interactions occurring on the hydrocarbon in the first FGD.

Each event in the dataset includes the energy-momentum four-vector for the leading muon and, if it exists, the leading proton in both truth and reconstructed space. For events that have been reconstructed, a detector sample ID is also included based on the sub-detectors used to measure the muon candidate (and proton candidate, if any), defined as one of:
\begin{enumerate}
    \item $\mu$TPC : only a reconstructed muon in the TPC
    \item $\mu$TPC + pTPC : 1 muon and 1 proton in the TPC
    \item $\mu$TPC + pFGD : 1 muon in the TPC and 1 proton in the FGD
    \item $\mu$FGD + pTPC : 1 muon in the FGD and 1 proton in the TPC
    \item $\mu$FGD : 1 muon in the FGD
    \item CC1$\pi$ : 1 muon and 1 $\pi^{+}$ in the TPC
    \item CCOther : 1 muon and multiple pion tracks in the TPC
    \item CCMichel :  1 muon and a michel tagged pion in the FGD   
\end{enumerate}
The final three sample IDs here are intended as background-enhanced samples and are used to help constrain backgrounds to the CC$0\pi$ analysis. For events in the truth space, a topology ID is included indicating what observable final state particles result from the interaction, defined as one of:
\begin{enumerate}\label{enum:topology}
    \item CC0$\pi$0$p$ : only an outgoing muon from the charged current interaction
    \item CC0$\pi$1$p$ : 1 muon and 1 proton
    \item CC0$\pi$N$p$ : 1 muon and 2 or more protons
    \item CC1$\pi$ : 1 muon and 1 charged pion
    \item CCOther : other events with a muon (neutral pion or multiple pion production)
\end{enumerate}

Events are designated as signal or background at truth level by these topology IDs, with 1--3 serving as signal and 4--5 serving as background. Due to limitations of the data release, this is not the full list of backgrounds typically included in a T2K analysis. In particular, it omits contributions from neutral current interactions, $\nu_e$ interactions with erroneous particle identification, wrong-sign events from $\bar{\nu}$ interactions, and out of fiducial volume interactions.

The dataset also contains information about systematic uncertainties associated with the flux, detector, and neutrino interaction models for the analysis. This comes in the form of a vector of 500 weight variations for each individual event that represent coherent random throws of these uncertainties. However, the detector systematic uncertainties do not include dedicated proton reconstruction effects, since the proton kinematics were not part of the original analysis associated with this data release.

\section{Test Setup}
\label{sec:TestSetup}
We create a set of ``fake data'' to serve as the unfolding target by reweighting a subset of the available events in the dataset with a modified charged-current quasi-elastic (CCQE) cross section based on the RPA calculation from Ref. \cite{NIEVES20161830}. This fake data represents a nontrivially different signal model from that used in the simulated dataset, which did not include this weak-charge screening effect caused by the nuclear medium. The modification is performed by applying an event weight to CCQE events as a function of the true $Q^2$ based on a parametrization of the RPA calculation using Bernstein polynomials, which is described in detail in Ref.~\cite{T2K:2021xwb}. The original set of non-reweighted events and weights then serve as the simulation for the unfolding procedure. Truth-level info for the fake data is not used during the unfolding procedures, but it is brought in for evaluation of the performance of each method in the final stage. For convenience, in the rest of this article we will refer to the fake data simply as ``data'', as during unfolding it is treated the same way that real data would be. Code used in the following analysis is publicly available \cite{OmnifoldGithub}.

For the events designated as simulation, we generate 500 throws of the event weights that simultaneously account for detector, cross-section, and flux systematic uncertainties, as well as Monte Carlo statistical uncertainties from the number of generated events. The weight variations from systematic effects are taken from the T2K public simulated data release. Monte Carlo statistical uncertainties are obtained by bootstrap resampling the available events with replacement. We similarly create 500 throws of the statistical uncertainties in the observed data by applying to it a weighted bootstrap resampling, where each data event with weight $w$ receives a resampled weight $w_{\mathrm{resampled}} = \mathrm{Poisson}(\lambda=w)$. This reduces the number of data events to the same statistics as Ref. \cite{T2K:PhysRevD.108.112009} and simulates the corresponding statistical uncertainty of that result. The unfolded results are then obtained by a toy universe method, where each of the 500 throws is independently run through the unfolding procedure. The resulting ensemble of 500 unfolded truth distributions gives our systematic and statistical uncertainties on the result. To maintain comparability between the unfolding methods under consideration, we use the same set of 500 variations for each of the procedures.

\begin{figure}
    \includegraphics[width=\linewidth]{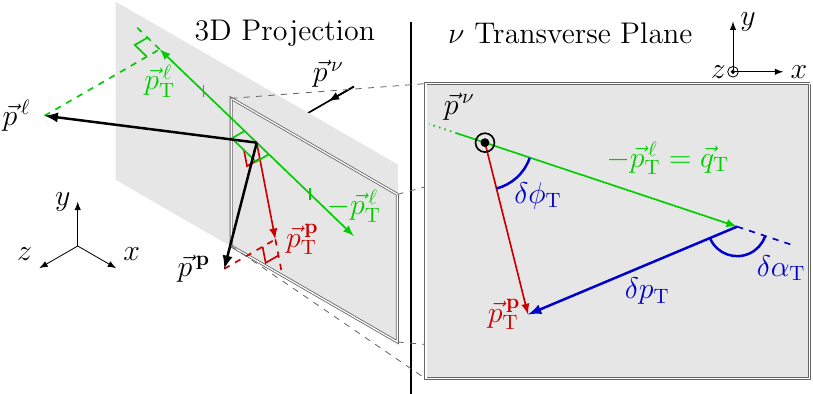}
    \caption{\label{fig:STVDiagram} Schematic view of the three single transverse variables for a quasi-elastic neutrino interaction producing a lepton with momentum $\vec{p}^{\,\ell}$ and a proton with momentum $\vec{p}^{\,p}$. The neutrino approaches from the $-z$ direction. On the right side, the variables are shown in the plane transverse to the initial neutrino momentum $\vec{p}^{\,\nu}$. Figure reproduced from Ref.~\cite{TKIProceedings}.}
\end{figure}

\begin{table}
\caption{\label{tab:MuonBinning} Binning of the muon momentum $p_{\mu}$ and forward angle $\theta_{\mu}$ used in extracting a binned differential cross section, identical to that used in the T2K analysis associated with the data release used in this work \cite{T2K:PhysRevD.108.112009}.}
\begin{tabular}{c|c|c}
    Bin Index  & cos $\theta_{\mu}$ & $p_{\mu}$ (GeV/c) \\ \hline
    0 & $-1.0 < \cos \theta_{\mu} < 0.2$ & $0 < p_{\mu} < 30$  \\ \hline
1 & \multirow{5}{*}{$0.2 < \cos \theta_{\mu} < 0.6$} & $0 < p_{\mu} < 0.3$  \\
2 &  & $0.3 < p_{\mu} < 0.4$  \\ 
3 &  & $0.4 < p_{\mu} < 0.5$  \\ 
4 &  & $0.5 < p_{\mu} < 0.6$  \\ 
5 &  & $0.6 < p_{\mu} < 30$  \\ \hline
6 & \multirow{6}{*}{$0.6 < \cos \theta_{\mu} < 0.7$} & $0 < p_{\mu} < 0.3$  \\ 
7 &  & $0.3 < p_{\mu} < 0.4$  \\ 
8 &  & $0.4 < p_{\mu} < 0.5$  \\ 
9 &  & $0.5 < p_{\mu} < 0.6$  \\ 
10 &  & $0.6 < p_{\mu} < 0.8$  \\ 
11 &  & $0.8 < p_{\mu} < 30$  \\ \hline
12 & \multirow{6}{*}{$0.7 < \cos \theta_{\mu} < 0.8$} & $0 < p_{\mu} < 0.3$  \\ 
13 &  & $0.3 < p_{\mu} < 0.4$  \\ 
14 &  & $0.4 < p_{\mu} < 0.5$  \\ 
15 &  & $0.5 < p_{\mu} < 0.6$  \\ 
16 &  & $0.6 < p_{\mu} < 0.8$  \\ 
17 &  & $0.8 < p_{\mu} < 30$  \\ \hline
18 & \multirow{7}{*}{$0.8 < \cos \theta_{\mu} < 0.85$} & $0 < p_{\mu} < 0.3$  \\ 
19 &  & $0.3 < p_{\mu} < 0.4$  \\ 
20 &  & $0.4 < p_{\mu} < 0.5$  \\ 
21 &  & $0.5 < p_{\mu} < 0.6$  \\ 
22 &  & $0.6 < p_{\mu} < 0.8$  \\ 
23 &  & $0.8 < p_{\mu} < 1.0$  \\ 
24 &  & $1.0 < p_{\mu} < 30$  \\ \hline
25 & \multirow{8}{*}{$0.85 < \cos \theta_{\mu} < 0.9$} & $0 < p_{\mu} < 0.3$  \\ 
26 &  & $0.3 < p_{\mu} < 0.4$  \\ 
27 &  & $0.4 < p_{\mu} < 0.5$  \\ 
28 &  & $0.5 < p_{\mu} < 0.6$  \\ 
29 &  & $0.6 < p_{\mu} < 0.8$  \\ 
30 &  & $0.8 < p_{\mu} < 1.0$  \\ 
31 &  & $1.0 < p_{\mu} < 1.5$  \\ 
32 &  & $1.5 < p_{\mu} < 30$  \\ \hline
33 & \multirow{7}{*}{$0.9 < \cos \theta_{\mu} < 0.94$} & $0 < p_{\mu} < 0.4$  \\ 
34 &  & $0.4 < p_{\mu} < 0.5$  \\ 
35 &  & $0.5 < p_{\mu} < 0.6$  \\ 
36 & & $0.6 < p_{\mu} < 0.8$  \\ 
37 & & $0.8 < p_{\mu} < 1.25$  \\ 
38 &  & $1.25 < p_{\mu} < 2.0$  \\ 
39 & & $2.0 < p_{\mu} < 30$  \\ \hline
40 & \multirow{10}{*}{$0.94 < \cos \theta_{\mu} < 0.98$} & $0 < p_{\mu} < 0.4$  \\ 
41 &  & $0.4 < p_{\mu} < 0.5$  \\ 
42 & & $0.5 < p_{\mu} < 0.6$  \\ 
43 &  & $0.6 < p_{\mu} < 0.8$  \\ 
44 & & $0.8 < p_{\mu} < 1.0$  \\ 
45 & & $1.0 < p_{\mu} < 1.25$  \\ 
46 &  & $1.25 < p_{\mu} < 1.5$  \\ 
47 &  & $1.5 < p_{\mu} < 2.0$  \\ 
48 &  & $2.0 < p_{\mu} < 3.0$  \\ 
49 & & $3.0 < p_{\mu} < 30$  \\ \hline
50 & \multirow{8}{*}{$0.98 < \cos \theta_{\mu} < 1.00$} & $0 < p_{\mu} < 0.5$  \\ 
51 &  & $0.5 < p_{\mu} < 0.7$  \\ 
52 & & $0.7 < p_{\mu} < 0.9$  \\ 
53 &  & $0.9 < p_{\mu} < 1.25$  \\ 
54 &  & $1.25 < p_{\mu} < 2.0$  \\ 
55 & & $2.0 < p_{\mu} < 3.0$  \\ 
56 &  & $3.0 < p_{\mu} < 5.0$  \\ 
57 & & $5.0 < p_{\mu} < 30$  \\

\end{tabular}
\end{table}

\begin{table}
\caption{\label{tab:STVBinning} Binning of the STVs $\delta p_\mathrm{T}, \delta \alpha_\mathrm{T}, \delta \phi_\mathrm{T}$ for analysis, copied from a previous T2K analysis that extracted differential cross sections as functions of these observables \cite{T2K:PhysRevD.98.032003}.}
\begin{tabular}{c|c|c}
$\delta p_\mathrm{T}$ (GeV) & $\delta \phi_\mathrm{T}$ (radians) & $\delta \alpha_\mathrm{T}$ (radians) \\
\hline
0.0-0.08 & 0.0--0.067 & 0.0--0.47\\
0.08-0.12 & 0.067--0.14 & 0.47--1.02\\
0.12-0.155 & 0.14--0.225 & 1.02--1.54\\
0.155-0.2 & 0.225--0.34 & 1.54--1.98\\
0.2-0.26 & 0.34--0.52 & 1.98--2.34\\
0.26-0.36 & 0.52--0.85 & 2.34--2.64\\
0.36-0.51 & 0.85--1.50 & 2.64--2.89\\
0.51-1.1 & 1.50--$\pi$ & 2.89--$\pi$\\
\end{tabular}
\end{table}

We consider four observables for evaluating the performance of each method:
\begin{itemize}
	\item ($p_{\mu}$, cos $\theta_{\mu}$) : the two-dimensional muon kinematics defined by its momentum and forward angle, which were  used in the differential cross-section analysis associated with the original data release \cite{T2K:PhysRevD.108.112009}
	\item $\delta p_\mathrm{T} \equiv |\vec{p}_{\mathrm{T}}^{\,\mu} + \vec{p}_{\mathrm{T}}^{\,p}|$ : the transverse momentum imbalance between the muon and leading proton
	\item $\delta \alpha_\mathrm{T} \equiv \arccos{\dfrac{-\vec{p}_{\mathrm{T}}^{\,\mu}\cdot\delta\vec{p}_{\mathrm{T}}}{p_{\mathrm{T}}^{\mu}\delta p_{\mathrm{T}}}}$ : the boosting angle between the muon and leading proton in the transverse plane
	\item $\delta \phi_\mathrm{T} \equiv \arccos{\dfrac{-\vec{p}_{\mathrm{T}}^{\,\mu}\cdot\vec{p}_{\mathrm{T}}^{\,p}}{p_{\mathrm{T}}^{\mu} p_{\mathrm{T}}^{p}}}$ : the angular difference between the muon and leading proton in the transverse plane compared to the case where they are back-to-back
\end{itemize}
In the definitions of the latter 3 variables, $\vec{p}_{\mathrm{T}}^{\,\mu}$ and $\vec{p}_{\mathrm{T}}^{\,p}$ are, respectively, the projections of the momenta of the muon and leading proton into the plane transverse to the incident neutrino direction. These variables are sometimes referred to as single transverse variables (STVs) and have been used to measure transverse kinematic imbalances in neutrino-nucleus interactions \cite{TKIIntro,T2K:PhysRevD.98.032003}. A visualization of the physical meaning of the STVs is provided in Fig. \ref{fig:STVDiagram}. The STVs are not calculated for events where there is no leading proton, which can happen either because the proton was not reconstructed or because no proton was kicked out at truth level.

While OmniFold is an unbinned procedure, we must choose some binning of the variables for final evaluation and comparison against classical binned techniques. The two-dimensional binning of the muon kinematics is summarized in Tab. \ref{tab:MuonBinning}, and the binning of the STVs is summarized in Tab. \ref{tab:STVBinning}. These binnings were chosen in previous T2K analyses \cite{T2K:PhysRevD.108.112009,T2K:PhysRevD.98.032003} based on the detector resolution and efficiency within the relevant kinematic phase spaces, as well as the expected statistics. For each of the binnings, we extract a differential cross section from the unfolded result and compare against the true cross sections used to generate the data.

In the OmniFold approach, we have the freedom to choose all of our inputs independently of the observables we wish to inspect in the end. We define three variations that we refer to as UniFold, MultiFold, and OmniFold in the same way as Ref.~\cite{Andreassen:2019cjw}, distinguished by which observables are provided as inputs in each case. UniFold is effectively an unbinned version of IBU, with the only kinematic input being a single observable we are trying to unfold. Correspondingly, we run four separate instances of UniFold: one for each of the observables used in evaluation of the unfolding performance. MultiFold takes as input $(p_{\mu}, \cos \theta_{\mu}, p_p, \delta p_\mathrm{T}, \delta \alpha_\mathrm{T}, \delta \phi_\mathrm{T})$, which directly includes all of the observables that we are interested in and allows for the simultaneous unfolding of all of them. OmniFold has the most general inputs\footnote{These are the most general inputs given what is provided in the available dataset, but a more detailed dataset would allow the use of an even wider phase space, potentially including all true/reconstructed particles.}, using just the kinematics of the muon and leading proton $(p_{\mu}, \cos \theta_{\mu}, \phi_{\mu}, p_p, \cos \theta_p, \phi_p)$, where, in the coordinates of Fig. \ref{fig:STVDiagram}, $\phi \equiv \arctan{(p_y / p_x)}$ is the direction of the particle in the plane transverse to the neutrino beam direction. The STVs can be derived from these inputs, but we can expect they would be more difficult to learn compared to MultiFold. We refer to these three variations collectively as the OmniFold method or approach, since they all operate on the same principles.

During training and evaluation of the classifiers used in the OmniFold method, the momentum variables are log-transformed to mitigate the effects of long tails and then standardized to a mean of 0 and standard deviation of 1. In events where one of the input variables is missing, either because it is missing at truth level or because of a failed reconstruction, a value of 0 is used for the missing quantity. For any particular setup, we use the same kinematic variables in both step 1 (reweighting of detector-level events) and step 2 (reweighting of truth-level events) of the OmniFold procedure, although this is not necessary in general~\cite{H1:2023fzk}. In addition, in step 1 for each case we also include the detector sample information, and in step 2 we include the event's truth-level interaction topology information, both of which are described in Sec. \ref{sec:Dataset}. These labels are one-hot encoded for input\footnote{One-hot encoding is a common way to deal with categorical information in neural networks. An input with N possible values is structured as N separate inputs $x_1, ..., x_N$. An event that belongs to category $i \subset \{1..N\}$ is given a value of 1 for input $x_i$, while all other inputs $x$ receive a value of 0 for that event.}. 

The OmniFold method is formally agnostic to the specific classifier architecture; for this work, we use a dense neural network with 4 hidden layers of 100 nodes each, using LeakyReLU activation functions for the hidden nodes and a sigmoid activation for the final output. The input features for each event are structured as a simple vector and received by an input layer with the corresponding number of nodes. The dataset that is the weighting target is given a label of 1, and the dataset that is to be reweighted is given a label of 0. In the case of step 1, these are respectively the data and the reconstructed simulation, and in the case of step 2, these are the generator events with pulled weights and the generator events with original weights\footnote{Note that the choice of labels is arbitrary. Flipping the labels would simply mean flipping the expression for the reweighting factor that comes from the classifier output.}. We train with a weighted binary cross-entropy loss, so that with an event from the 0-labeled dataset with a classifier result $0<q<1$ will receive a reweighting factor of $q/(1-q)$. This reweighting factor can be capped to some maximum value for events where $q$ is close to 1, but we did not find this to be necessary in our results. We note that obtaining extreme values of $q$ from the classifier would imply some kind of severe disagreement between data and simulation, which would generally indicate the need to reconsider the choice of unfolding inputs or phase space.

The network is implemented with Keras \cite{Keras} in Tensorflow v2.9.0 \cite{Tensorflow} and trained with the Adam optimizer \cite{Adam}. We use an initial learning rate of $1\times 10^{-4}$, a batch size of 1024, and a 80/20 training/validation split of the available events. Training continues until there has been no improvement in the validation loss for 15 epochs, after which the model weights that gave the best validation loss are restored as the final result for that iteration. Training of a single network generally took less than 30 seconds on a NVIDIA A100 GPU. On subsequent iterations past the first, the same total set of available events is used, but with a different random training/validation split each time. These later iterations also usually take less time to train, as they are warm-started using the weights from the previous iterations.

\begin{figure}
    \includegraphics[width=\linewidth]{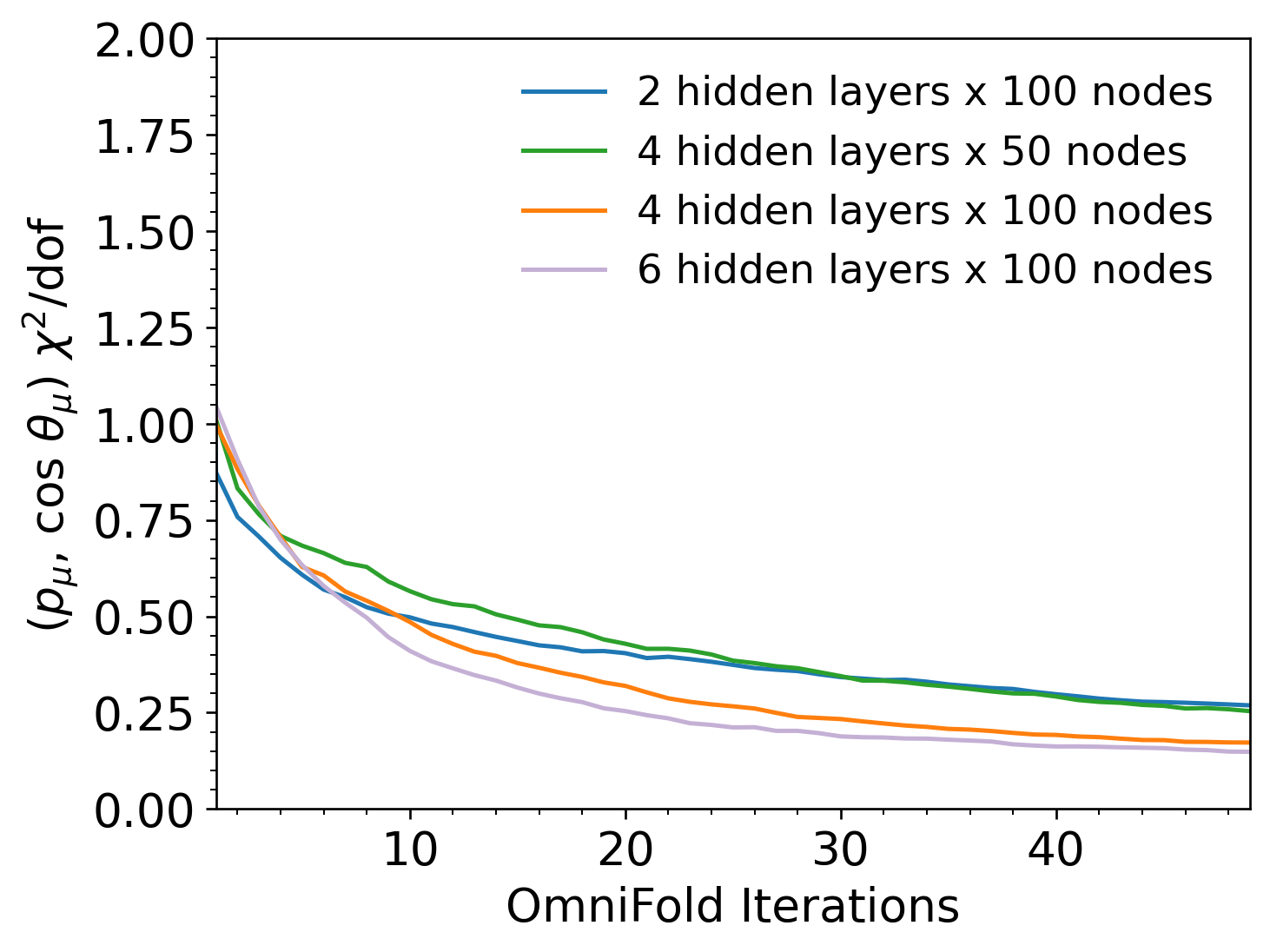}
    \caption{\label{fig:NNSizeComparison} Comparison of $\chi^2$ performance for the unfolded differential cross section as a function of ($p_{\mu}, \cos \theta_{\mu}$) with OmniFold using different sizes for the neural networks that serve as the classifiers in the method. Networks that are too small lack the ability to fully learn the high-dimensional likelihood ratio and result in worse unfolding performance, but performance also flattens out once additional network capacity becomes extraneous.}
\end{figure}

We performed a coarse scan of the training parameters and did not find significant improvement from other values. However, we did find that an insufficiently large neural network size resulted in notable degradation of the performance, as one would expect when the network loses the capability to properly learn the likelihood ratio between the data and simulated distributions. A comparison of the performance with different network sizes is shown in Fig. \ref{fig:NNSizeComparison}. The final choice of 4 hidden layers with 100 nodes each was chosen as a point beyond which expanding the network size further had no obvious benefit.

To serve as a benchmark of a typical conventional unfolding approach, we perform an unfolding of each observable of interest with a binned version of the OmniFold method, which with perfect training is mathematically equivalent to IBU \cite{Andreassen:2019cjw, BDTUnfold}. In this setup, which we call IBU-UniFold, we follow the OmniFold reweighting procedure, but for each event we provide only the bin index for the observable in question instead of actual kinematic information. The bin index is provided as one-hot encoded input, in addition to one-hot encoded information about the sample ID in step 1 and topology ID in step 2 like what was provided in the other OmniFold methods. The use of IBU-UniFold allows us to have a background treatment equivalent to what is used in the OmniFold methods, which makes direct comparison of the performance between methods more straightforward. This is in contrast with the background subtraction techniques typically used for IBU, which are different enough that the results and uncertainties are complicated to compare against those from OmniFold.

As an additional benchmark, we also run a binned variation that we call Binned UniFold. This differs from IBU-UniFold only in that we provide the bin information via the value of the bin center in the corresponding kinematic phase space instead of as one-hot encoded bin indices. In principle, there is no change in the amount of provided information between the Binned UniFold setup and the IBU-UniFold setup. However, providing bin centers results in the classifier needing to smoothly interpolate between adjacent bins in the phase space, which can make training more difficult compared to the one-hot encoded case.

Background subtraction and efficiency corrections do not require explicit treatment with the OmniFold method. Instead, these are implicitly~\cite{Andreassen:2021zzk} applied when we obtain the reweighted generator events, which can be individually identified as being signal or background and passing selection cuts or not. Different options are available for technical details of how to handle selection cuts during the OmniFold procedure, the impact of which we discuss further in Sec.~\ref{sec:results}. For evaluation of the performance, all results are converted into differential cross sections using the same bins.

We run 50 iterations of each of the unfolding procedures, so that their convergence rates can be compared and a smaller cutoff on the number of iterations can be imposed later to serve as the regularized final result. In order to mitigate uncertainties from the stochastic nature of the neural network training process, we run 5 trials of each OmniFold procedure using different random training/test splits of the events shown to the neural network each time. The reweighting factors from the resulting networks are then averaged to obtain the unfolded result.

\section{Performance and Results}
\label{sec:results}
\subsection{Convergence Criteria}

To evaluate performance, we calculate the CC0$\pi$ (signal) differential cross sections in the chosen kinematic bins. For the single transverse variables, we impose an additional requirement that the proton have a minimum momentum of 450 MeV at truth level, which is roughly the proton reconstruction threshold in the ND280 detector \cite{T2K:PhysRevD.98.032003}. We calculate a metric $\chi^2 = (\vec{\mu} - \vec{x})^\intercal \Sigma^{-1} (\vec{\mu} - \vec{x})$ for each of the binned differential cross sections in truth space, using the unfolded result obtained from each of the methods. The vectors $\vec{\mu}$ are the mean binned results for the observable in question over the 500 simulated pseudo-experiments with the previously described systematic and statistical variations, and the covariance matrix $\Sigma$ is obtained from the spread of results over those same pseudo-experiments. The vectors $\vec{x}$ are the truth-level differential cross sections for the data that we have created as the unfolding target.

\begin{figure}
    \includegraphics[width=\linewidth]{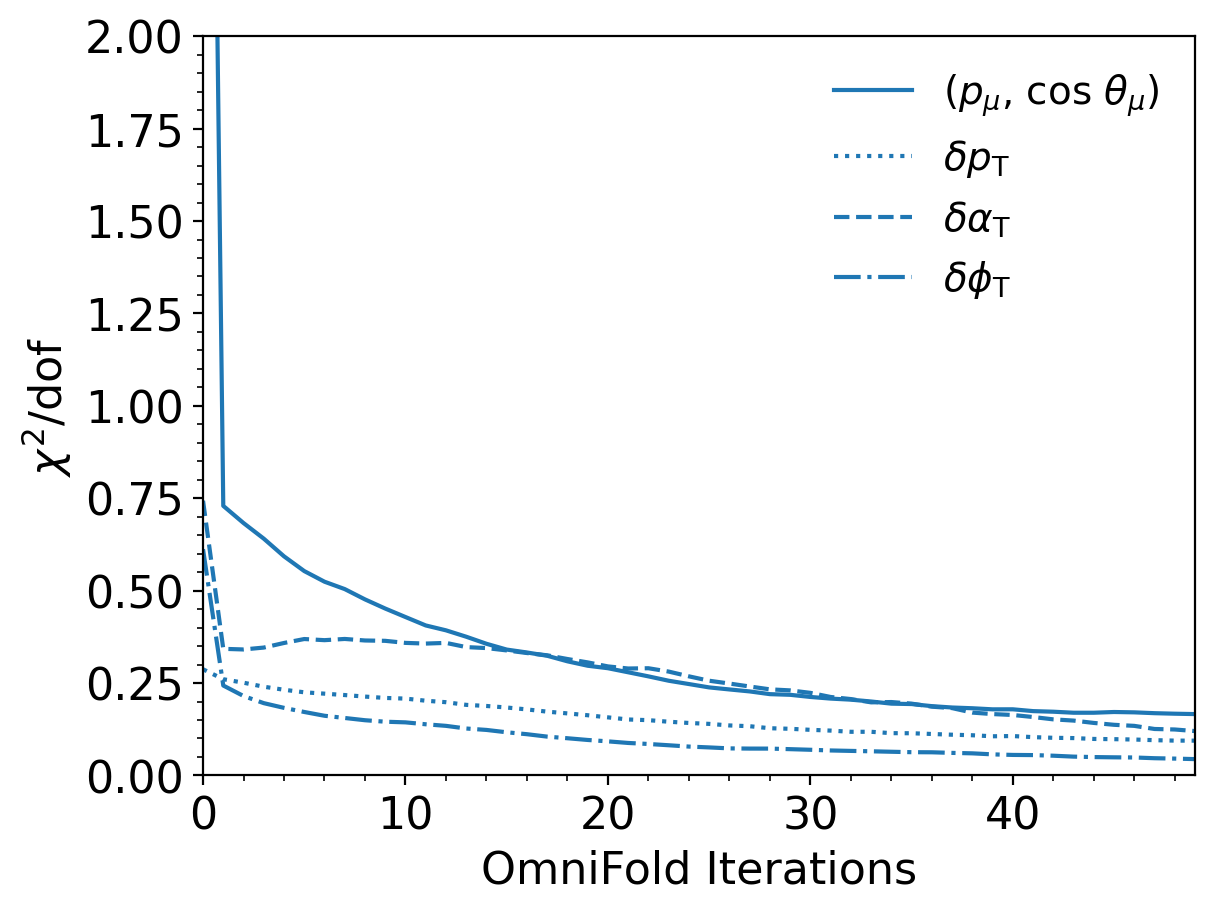}
    \caption{\label{fig:Chi2Iterations} $\chi^2$/DoF for the OmniFold result as a function of number of iterations of the procedure, for each of the differential cross sections under consideration. The 0th iteration is the prior, which begins outside the y-axis limits for $(p_{\mu}, \cos \theta_{\mu})$. As with any iterative procedure, using a larger number of iterations reduces the dependence on the prior and reduces the amount of bias in the result. The point at which the $\chi^2$ appears to converge depends on the observable and binning under inspection.}
\end{figure}

The $\chi^2$ comparison of the OmniFold results against the truth-level data as a function of number of iterations of the unfolding procedure is shown in Fig. \ref{fig:Chi2Iterations}. For the purposes of this study where we wish to compare the bias and uncertainties between methods, we choose an iteration number at which the $\chi^2$ curve has flattened out to be considered the converged result, leaning towards a larger number of iterations to minimize bias. We choose a cutoff of 20 iterations for IBU-UniFold and 45 iterations for the OmniFold methods. However, this is not a feasible approach for a real data analysis, since it relies on knowing the truth-level target distribution. Furthermore, it relies on having a particular binning in mind already, which is potentially at odds with the general promise of the OmniFold method allowing for unbinned unfolding of multiple observables simultaneously.

\begin{figure}
    \includegraphics[width=\linewidth]{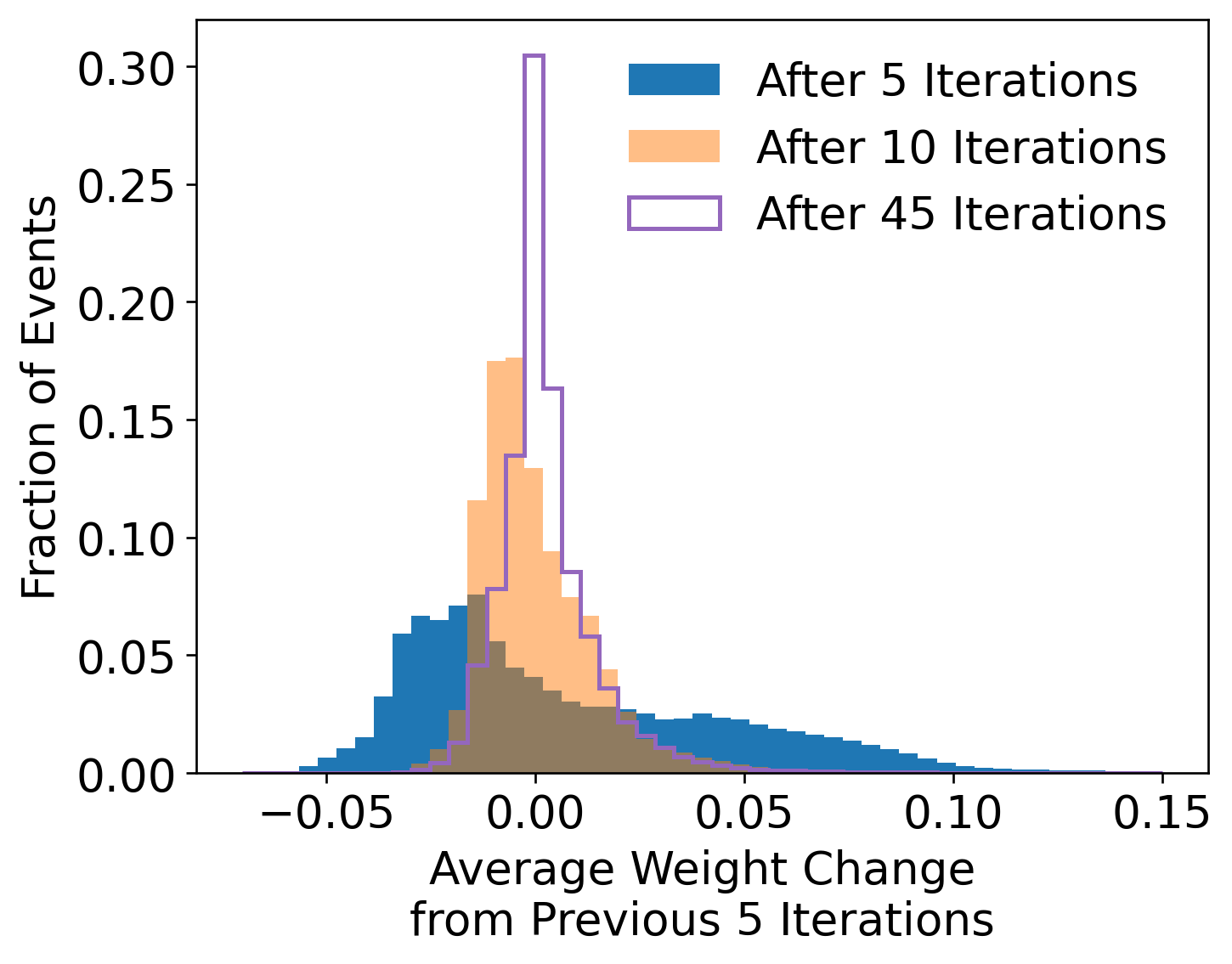}
    \caption{\label{fig:OmnifoldConvergence} The average weight change that each event has received over the previous 5 iterations of the OmniFold procedure, after 5, 10, and 45 total iterations. After 5 and 10 iterations, one can see the distributions of weight changes are still obviously skewed, indicating that there are significant reweightings still occurring across the whole dataset. After 45 iterations, we see the distribution is centered around 0 with not much of a tail. This suggests the unfolding procedure is mostly done at this point, which is also consistent with the behavior of the $\chi^2$ curves shown in Fig. \ref{fig:Chi2Iterations}.}
\end{figure}

As one possibility for an unbinned convergence criterion, we note that the OmniFold method functions by reweighting each individual event after each iteration. When the result is converged, running further iterations should only result in random jitter of the event weights instead of any significant changes. We propose considering the weight change per event averaged over the last $N$ iterations, for some choice of $N$. As the result converges, the distribution of these weight changes should approach a Gaussian centered at 0 with some characteristic width determined by the problem. A sample of how these distributions look for $N=5$ in our dataset is shown in Fig. \ref{fig:OmnifoldConvergence}. While here we have not imposed a quantitative metric for convergence from these distributions, qualitatively, it is clear that the distributions at larger numbers of iterations are closer to the expected distribution of a converged result than the ones at smaller numbers of iterations. Future work could explore this further as an avenue towards a principled way to choose the degree of regularization in an unbinned manner.

\begin{figure}
    \includegraphics[width=\linewidth]{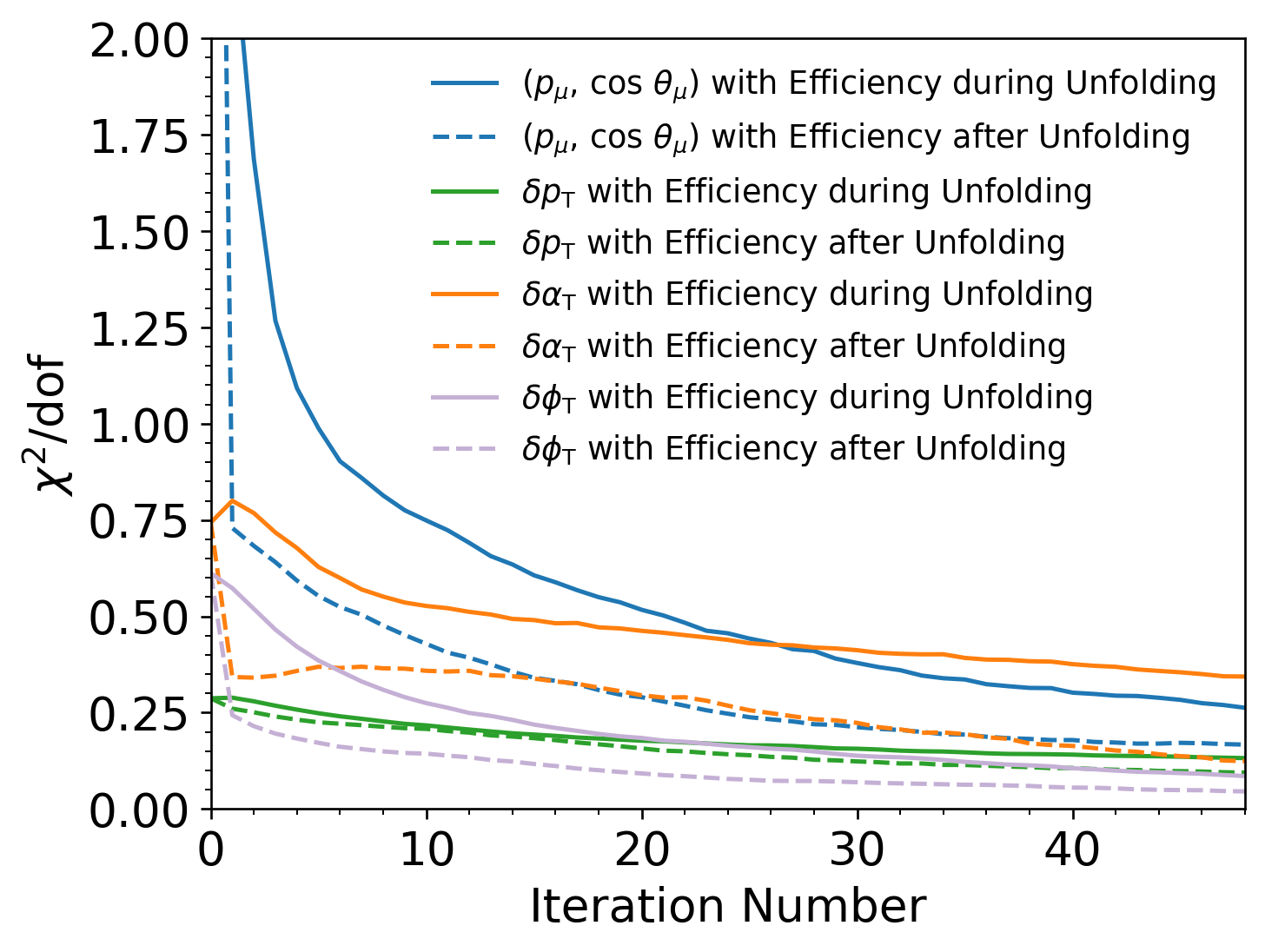}
    \caption{\label{fig:EfficiencyEffects} OmniFold $\chi^2$ performance for each differential cross section of interest as a function of number of iterations, using different efficiency treatments. Including efficiency effects during unfolding (solid lines) means including non-reconstructed events in step 2 of the OmniFold procedure, where truth-level event reweighting is done. Applying efficiency corrections after unfolding (dashed lines) means excluding non-reconstructed events from step 2 but applying reweighting factors to them when they are brought in for the final unfolded result, which we see yields faster and better convergence.}
\end{figure}

One more issue that impacts the convergence rate is the question of efficiency treatment with the OmniFold method, which is worth explicit discussion. Broadly speaking, there are two options: including efficiency effects during unfolding or applying them afterwards. In the first case, we would include truth events that have no reconstructed counterpart during the step 2 reweighting of OmniFold, simply applying either a pull weight of 1 or some other average weight for those events. In the second case, we would exclude these truth events with failed reconstructions from the iterative procedure completely, but reintroduce them when evaluating the final result. In this case, they would receive a reweighting factor that is an extrapolation by the classifier trained from the OmniFold procedure, and this is equivalent to applying an unbinned efficiency correction after the unfolding procedure. A comparison of the performance of these approaches for our setup is shown in Fig. \ref{fig:EfficiencyEffects}. We expect little difference in results between the two approaches when efficiencies are generally high for all bins, but this is not the case for efficiencies in our analysis. We observe that including efficiency effects during unfolding considerably slows the rate of $\chi^2$ convergence, since this increases the bias towards the prior in regions of low efficiency. Correspondingly, the rest of the results in this paper are all obtained using the approach of an efficiency correction after unfolding.

\subsection{Results}

The resulting $\chi^2$ values comparing the unfolded truth-level differential cross sections against those underlying the data are shown in Tab. \ref{tab:AllChi2} for each of the observables of interest and for each of the multiple unfolding approaches described in Sec. \ref{sec:TestSetup}. Since the $\chi^2$ metric does not clearly distinguish between results with large bias and large uncertainty and results with small bias and small uncertainty, we also consider the triangular discriminator $\Delta(P,Q) = \sum_{i} |p_i - q_i|^2 / (p_i + q_i)$~\cite{850703}, where $p_i$ is the unfolded content of bin $i$, $q_i$ is the true content of bin $i$, and the summation runs over the available bins. The results from evaluating the triangular discriminator between each unfolded result and the true data distribution are shown in Tab. \ref{tab:TriangularDiscrim}, serving as an estimate of solely the bias. We see that MultiFold and OmniFold are able to achieve $\chi^2$ values comparable to those of IBU-UniFold across all observables, while also achieving less bias. We emphasize that the IBU-UniFold and UniFold results are obtained by running a separate unfolding procedure for each different observable, whereas in the MultiFold and OmniFold results we are simultaneously unfolding all of them in one pass. 

\begin{table}
\caption{\label{tab:AllChi2} $\chi^2$ values from comparing unfolded differential cross sections using each method to the truth-level cross sections underlying the data. Results are divided by observable of interest.}
\begin{tabular}{c|c|c|c|c}
& \multicolumn{4}{c}{$\chi^2$} \\ \hline
Method & \makecell{($p_{\mu}$, cos $\theta_{\mu}$)\\DoF=58} & \makecell{$\delta p_{\mathrm{T}}$\\DoF=8} & \makecell{$\delta \alpha_{\mathrm{T}}$\\DoF=8} & \makecell{$\delta \phi_{\mathrm{T}}$\\DoF=8} \\ \hline
Prior & 298.2 & 2.3 & 5.9 & 4.9 \\ 
IBU-UniFold & 2.1 & 0.2 & 0.4 & 0.1 \\
Binned UniFold & 21.4 & 1.4 & 0.9 & 0.5 \\
UniFold & 27.1 & 1.1 & 0.6 & 1.1 \\ 
MultiFold & 3.1 & 0.3 & 0.2 & 0.3 \\ 
OmniFold & 10.0 & 0.8 & 1.1 & 0.4 \\ \hline
\end{tabular}
\end{table}

\begin{table}
\caption{\label{tab:TriangularDiscrim} Triangular discriminator values from comparing unfolded differential cross sections using each method to the truth-level cross sections underlying the data. Results are divided by observable of interest.}
\begin{tabular}{c|c|c|c|c}
& \multicolumn{4}{c}{Triangular Discriminator} \\ \hline
Method & ($p_{\mu}$, cos $\theta_{\mu}$) & $\delta p_{\mathrm{T}}$ & $\delta \alpha_{\mathrm{T}}$ & $\delta \phi_{\mathrm{T}}$ \\ \hline
Prior & 545.6 & 27.5 & 31.2 & 26.7 \\ 
IBU-UniFold & 17.1 & 1.9 & 3.4 & 0.8\\
Binned UniFold & 29.9 & 2.8 & 6.0 & 1.9 \\
UniFold & 17.3 & 5.7 & 5.8 & 1.7 \\ 
MultiFold & 2.7 & 0.7 & 0.6 & 0.6 \\ 
OmniFold & 9.4 & 1.7 & 3.0 & 2.1 \\ \hline
\end{tabular}
\end{table}

IBU-UniFold performs better than Binned UniFold, even though in principle these methods have access to the same amount of information. The discrepancy in performance between these methods is attributed to difficulties in training the neural networks, as learning the bin-to-bin fluctuations is easier with discrete inputs\footnote{Neural networks are known to be more effective in higher dimensions, which is the case for the one-hot-encoded entries.}. These training difficulties are likely exacerbated by the generally low statistics in a neutrino dataset compared to previous applications of OmniFold, with only around 20000 events in the data for our setup. Given this limitation, the similarity in performance between Binned UniFold and unbinned UniFold is unsurprising too, as our analysis bins are already quite fine relative to the available statistics.

\begin{figure}
    \includegraphics[width=\linewidth]{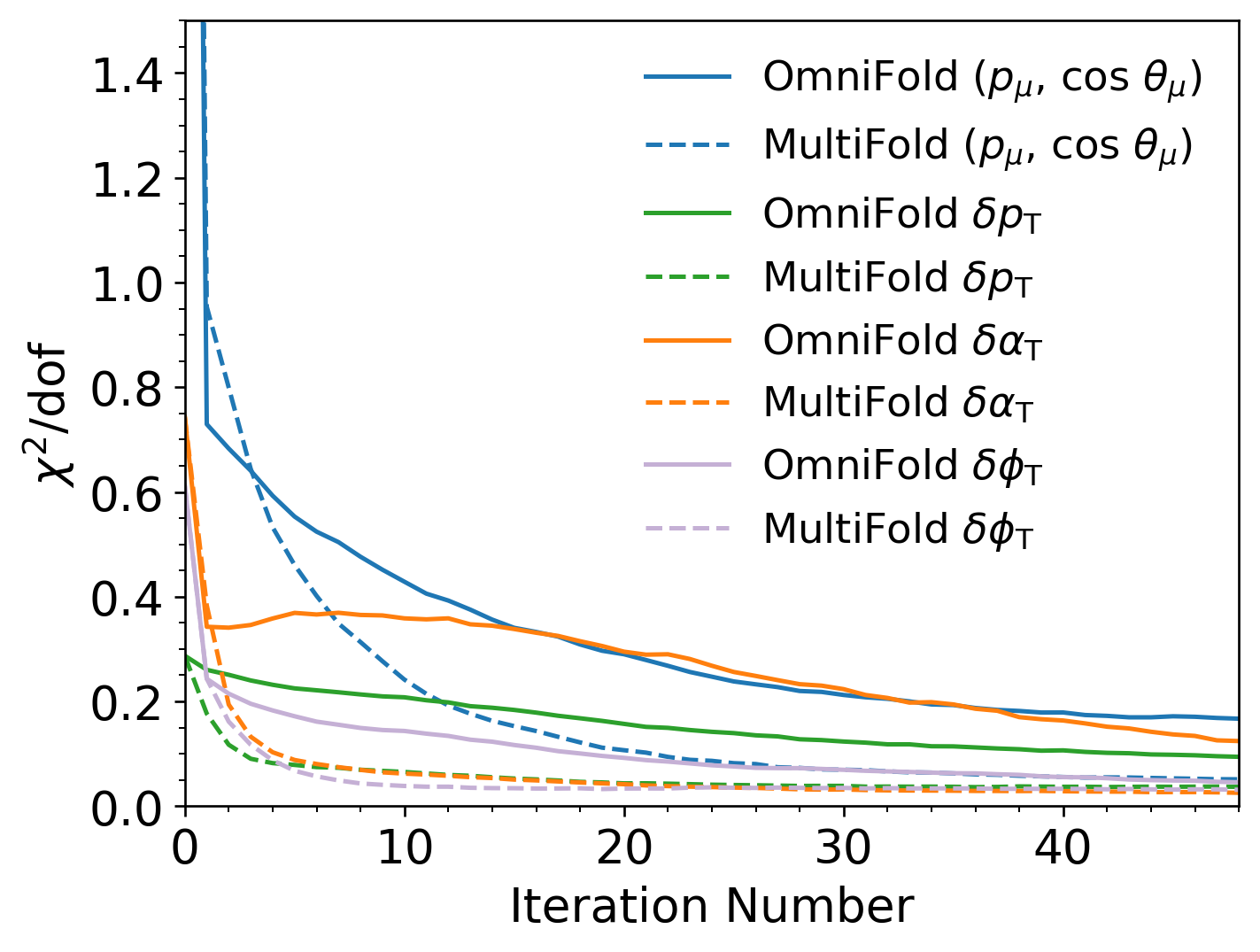}
    \caption{\label{fig:Inputcomparison} Comparison of $\chi^2$ performance for different inputs to the neural network in the OmniFold approach. For OmniFold (solid lines), the inputs are the muon and leading proton kinematics $(p_{\mu}, \cos \theta_{\mu}, \phi_{\mu}, p_p, \cos \theta_p, \phi_p)$, from which our observables of interest can be derived. With MultiFold (dashed lines), the kinematic variables used as input are $(p_{\mu}, \cos \theta_{\mu}, p_p, \delta p_{\mathrm{T}}, \delta \alpha_{\mathrm{T}}, \delta \phi_{\mathrm{T}})$, directly including the observables of interest.}
\end{figure}

The clear improvement in performance seen when comparing MultiFold and OmniFold against UniFold demonstrates that the additional information provided in those cases is indeed helpful in unfolding. The observation that MultiFold performs better than OmniFold across all metrics also serves as another indication that the classifier training is imperfect. A direct comparison of the $\chi^2$ convergence of these two methods is shown in Fig. \ref{fig:Inputcomparison}, where we see MultiFold converges both faster and to better values. While in principle OmniFold has access to all the information that MultiFold did in our setup, it is a nontrivial function to derive the STVs from the muon and proton kinematics. With the limited statistics in our dataset, it is unsurprising that the OmniFold networks appear to be unable to perform as well as the MultiFold networks, given that the latter directly receive all of the variables of interest. From a practical perspective, an analyzer who knows the observables they will be interested in will usually want to include this information as direct input to the procedure anyway, sidestepping this problem, but this calls for caution when attempting to unfold arbitrarily complicated derived quantities from simple inputs.

\begin{figure}
    \includegraphics[width=.49\linewidth]{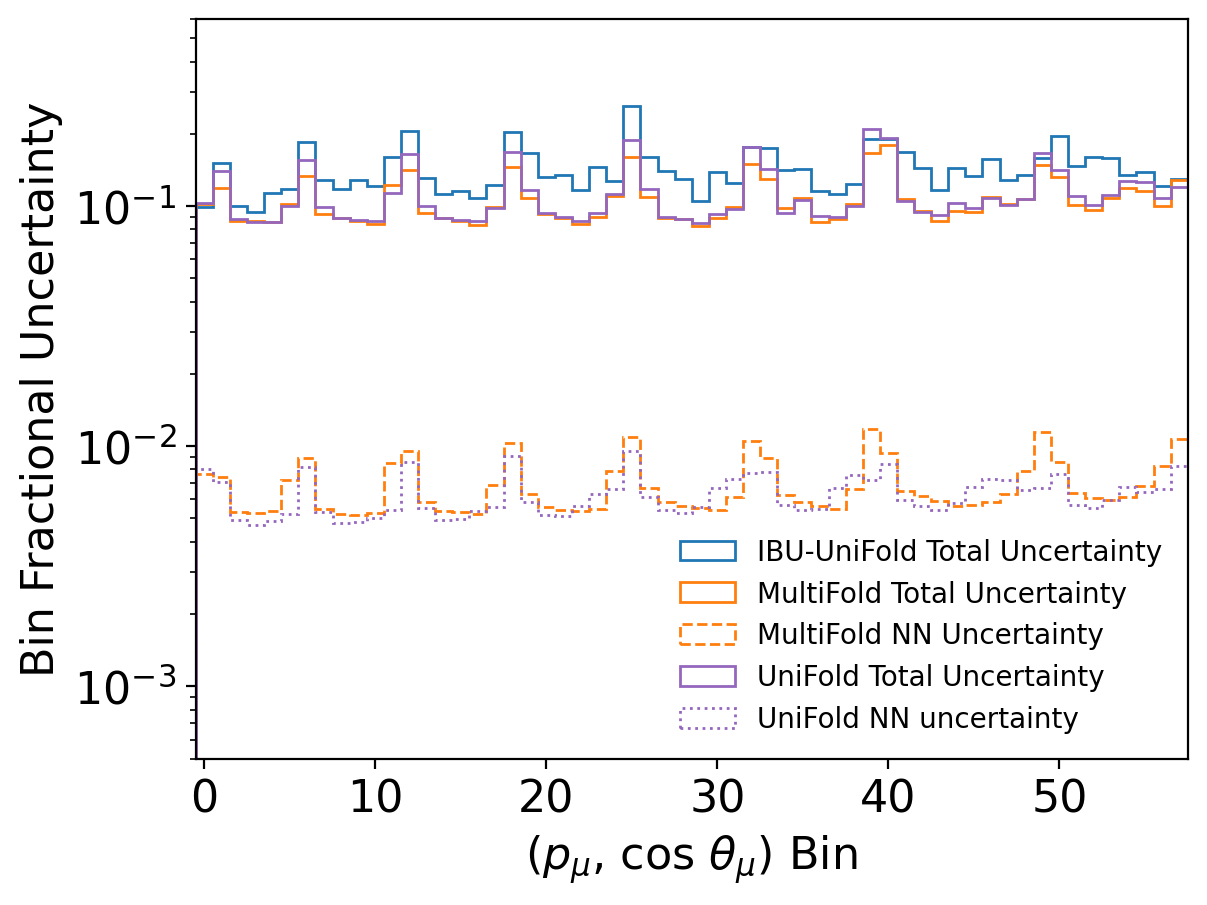}
    \includegraphics[width=.49\linewidth]{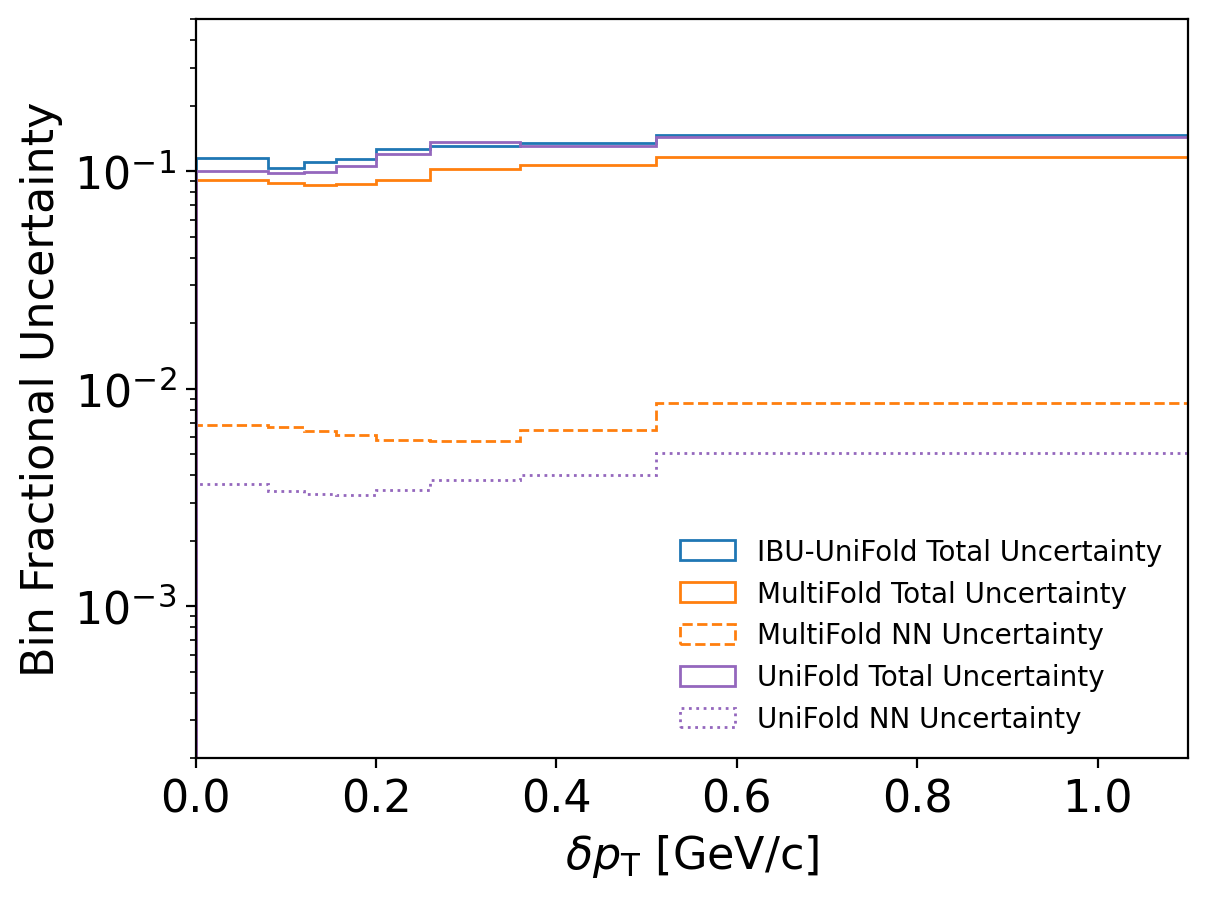}
    \includegraphics[width=.49\linewidth]{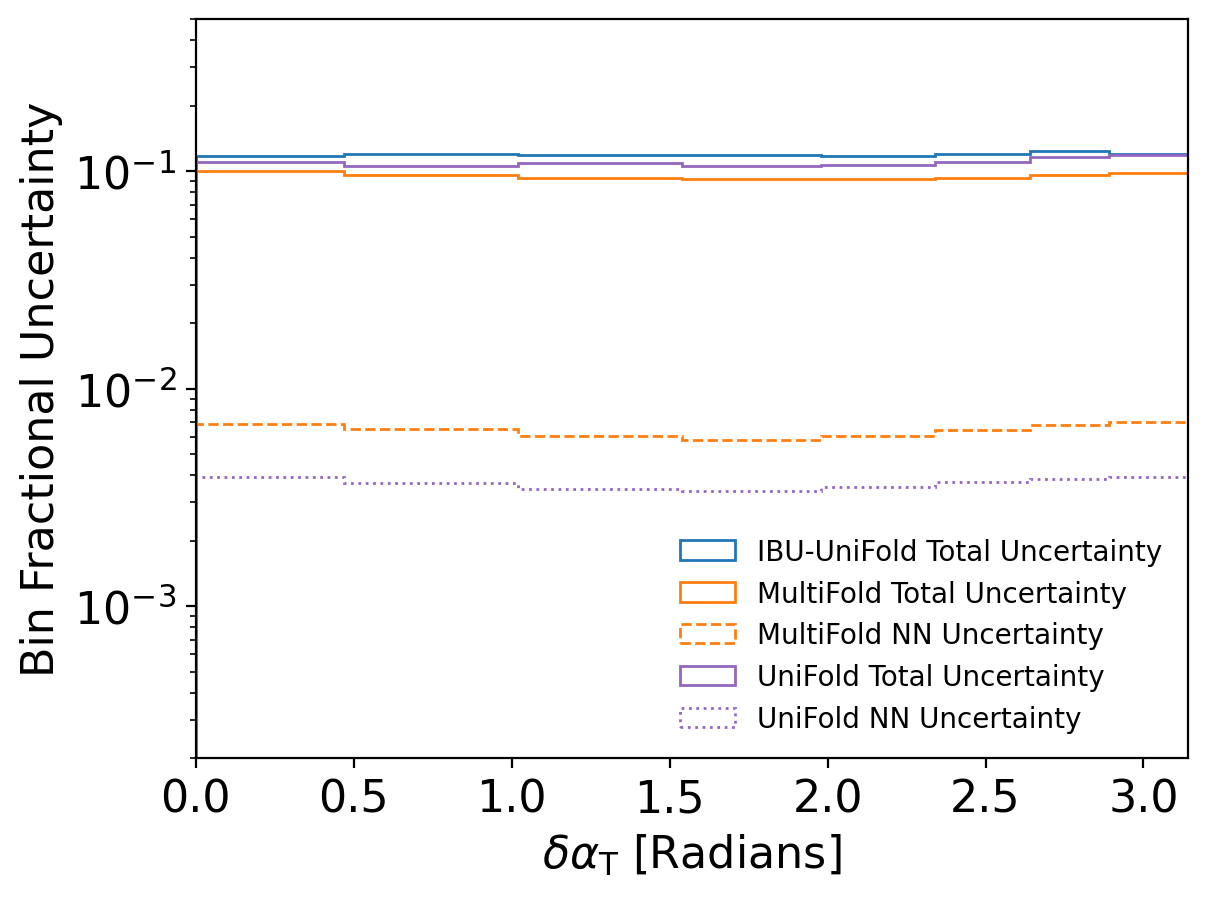}
    \includegraphics[width=.49\linewidth]{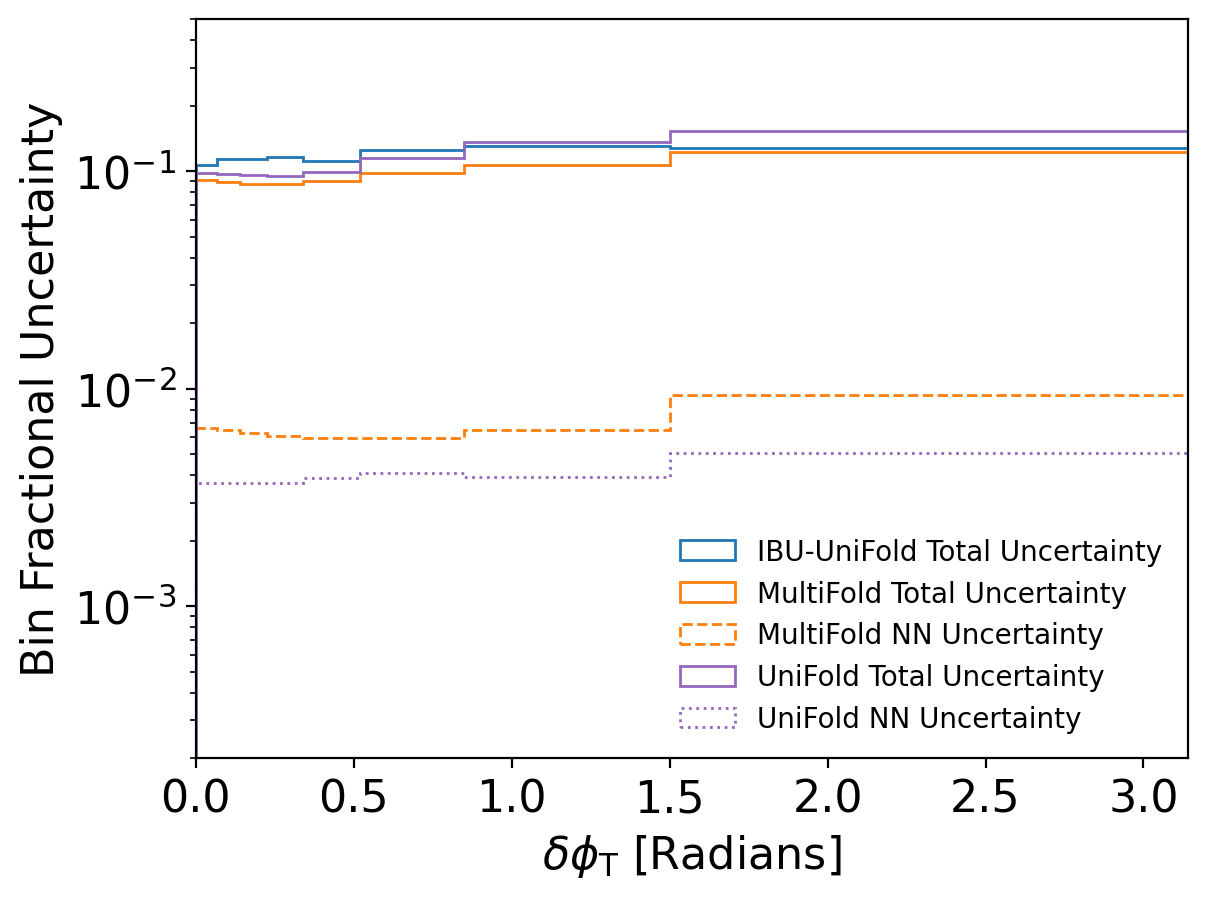}
    \caption{\label{fig:UncertaintyBudget} Bin-by-bin uncertainty in the unfolded results from IBU-UniFold, UniFold, and MultiFold, excluding effects of correlations among bins. The neural network (NN) uncertainty indicates the standard error due to just the neural network's stochastic training process, obtained by running the procedure 5 separate times. The MultiFold result contains generally lower uncertainties than IBU-UniFold. We also note that the NN uncertainty from UniFold is lower than that of MultiFold in the transverse variables, which is expected as a result of the simpler classifier problem that UniFold faces.}
\end{figure}

The fact that MultiFold achieves similar $\chi^2$ performance to IBU-UniFold for any particular observable is noteworthy given that it has the additional advantage that it is unfolding them all simultaneously. On top of this, the triangular discriminator performances indicate that MultiFold and OmniFold achieve less bias than IBU-UniFold. The discrepancy between the $\chi^2$ and triangular discriminator comparisons among methods can be attributed to the generally lower uncertainties obtained by the OmniFold methods. A comparison of the uncertainty budgets between IBU-UniFold, UniFold, and MultiFold is provided in Fig. \ref{fig:UncertaintyBudget}. We see that for the $(p_{\mu}, \cos \theta_{\mu})$ distribution in particular with its relatively fine bins, IBU-UniFold yields notably higher uncertainties. The unbinned nature of the OmniFold method's reweighting in combination with the penalization of overfitting by the training/validation split of the data results in a more regularized result compared to IBU-UniFold.

The uncertainty budget shown in Fig.~\ref{fig:UncertaintyBudget} also includes the contribution to the uncertainty from neural network training randomness, evaluated by calculating the variation in results using identical data but different training initializations for the neural networks. However, this standard error can be reduced by simply running the procedure additional times and averaging all of the results. The unfolded results we have presented are the result of ensembling 5 trials of neural networks, which reduces the standard error to a level that is negligible relative to the systematic and statistical uncertainties involved with the data. As expected given this small contribution to the uncertainty, we also find that the neural network ensembling has negligible impact on the overall unfolding performance. However, we note that the effect of ensembling will vary with the specific unfolding problem and classifier choice.

\begin{figure}
\centering
\includegraphics[width=0.3\linewidth]{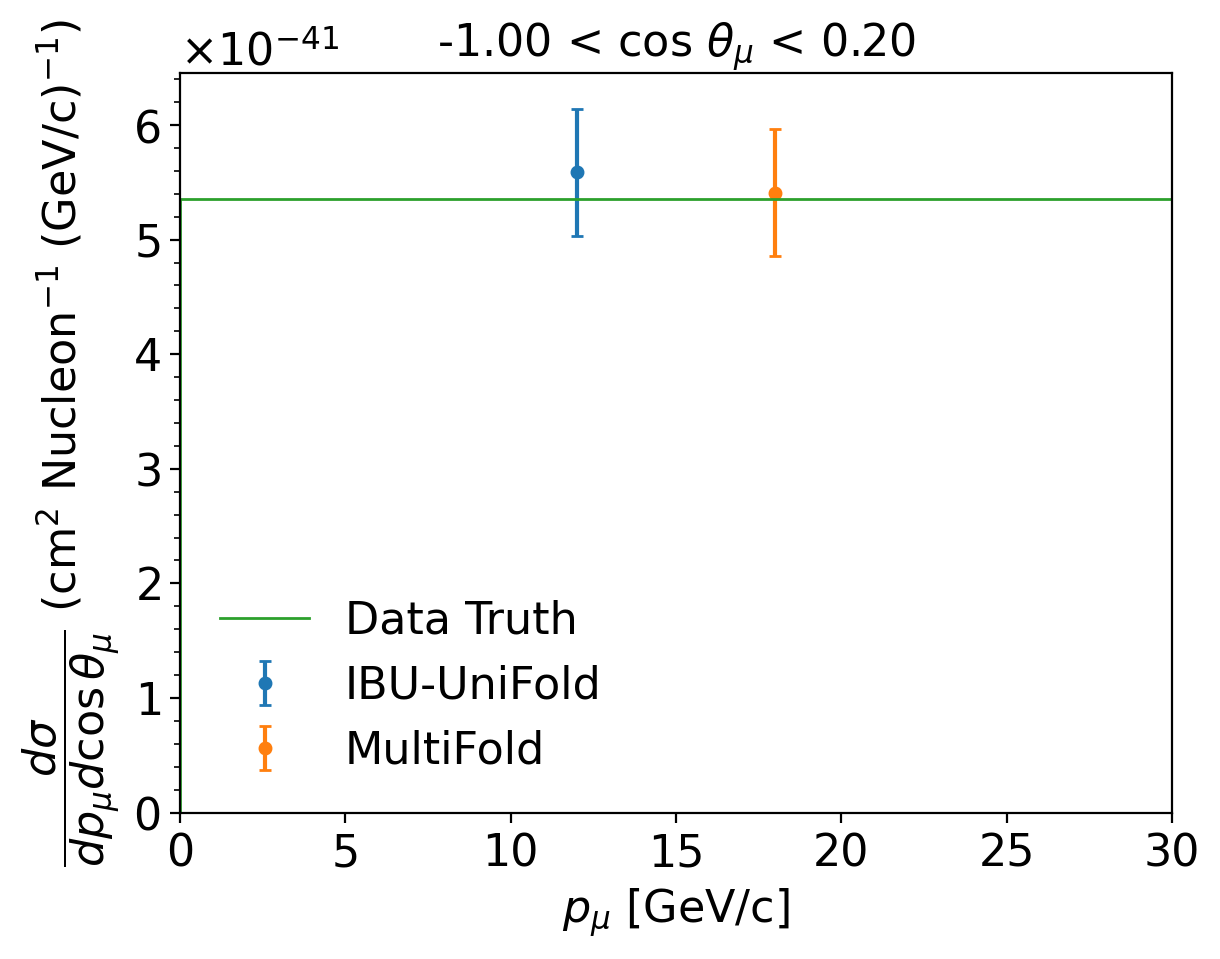}
\includegraphics[width=0.3\linewidth]{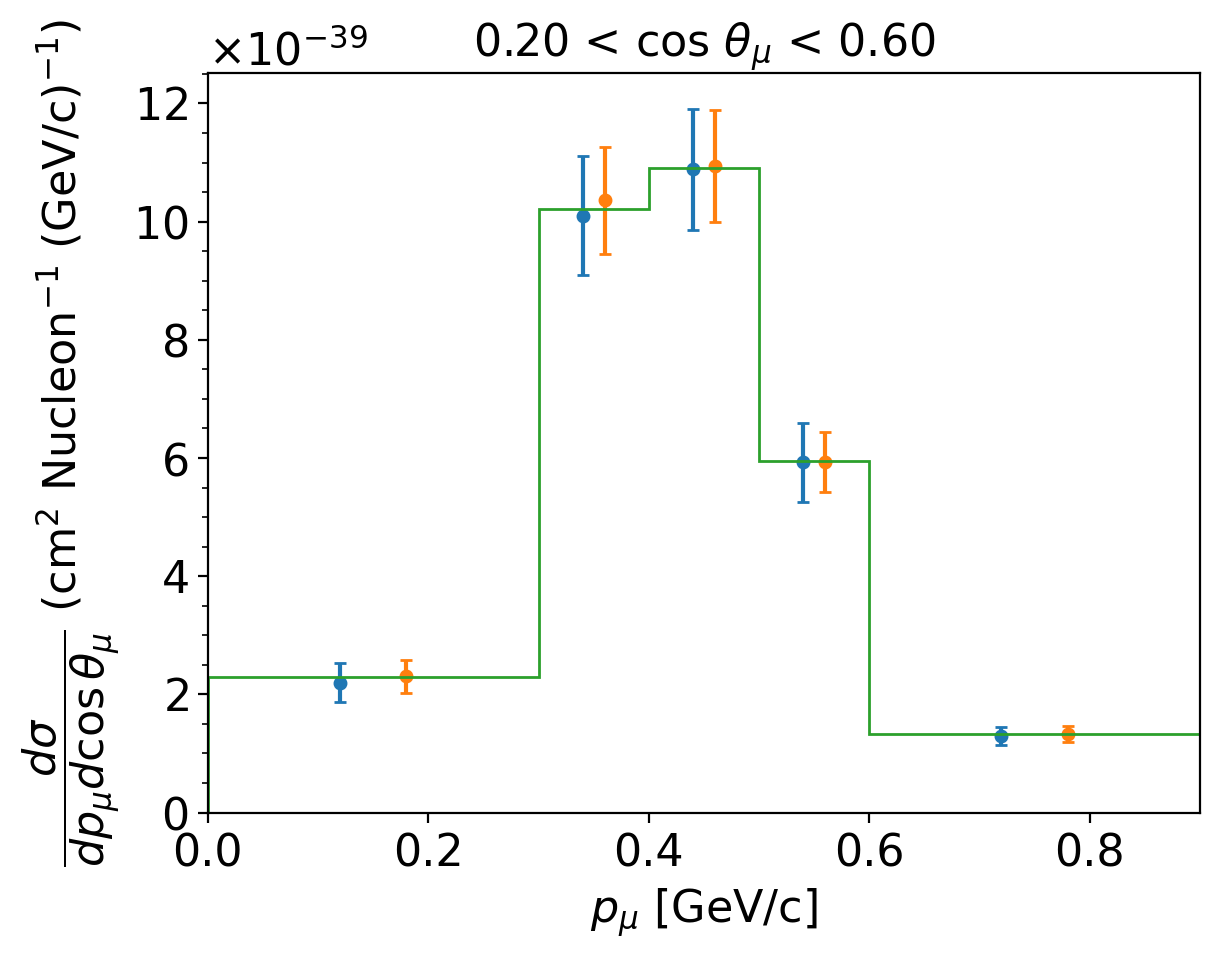}
\includegraphics[width=0.3\linewidth]{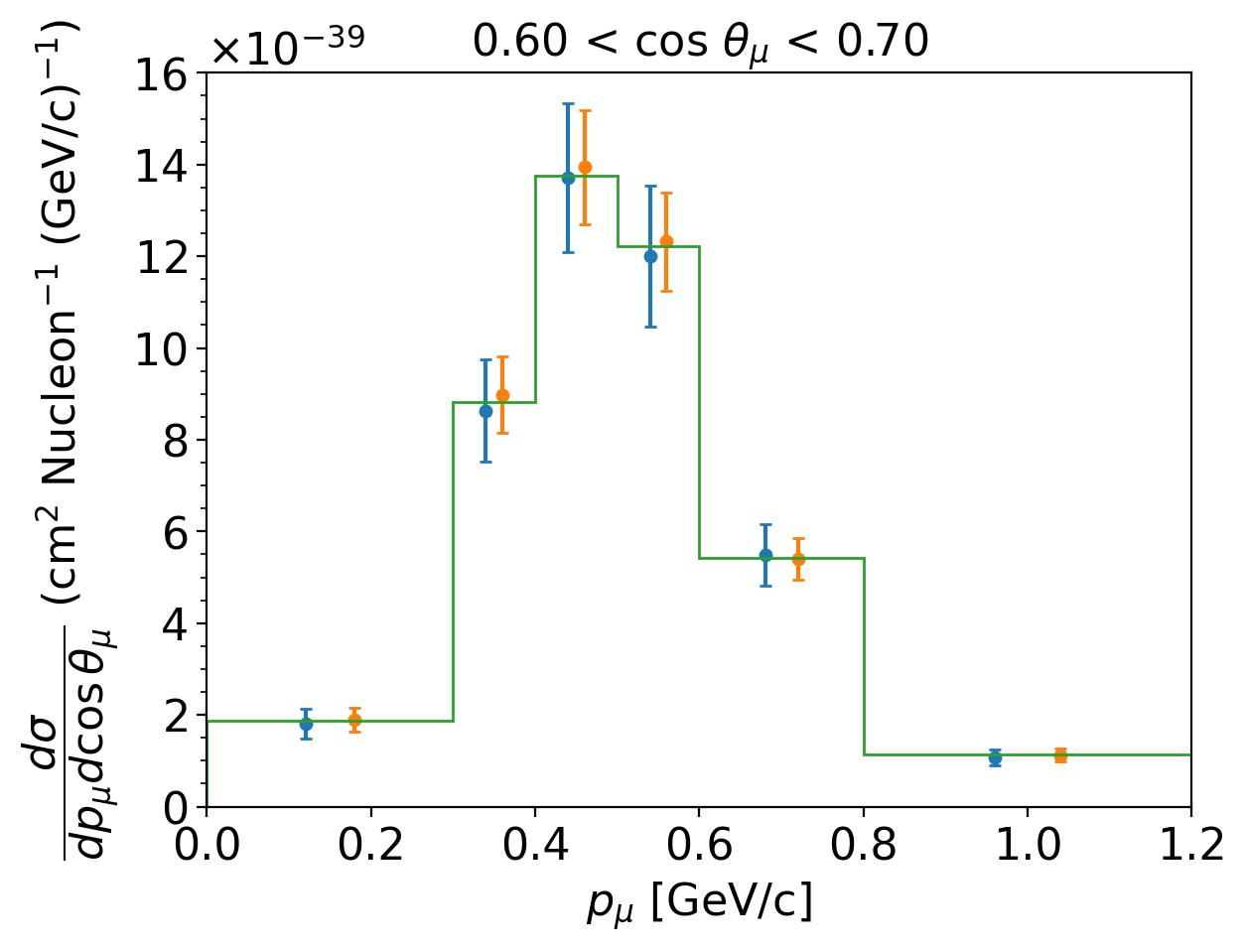}
\includegraphics[width=0.3\linewidth]{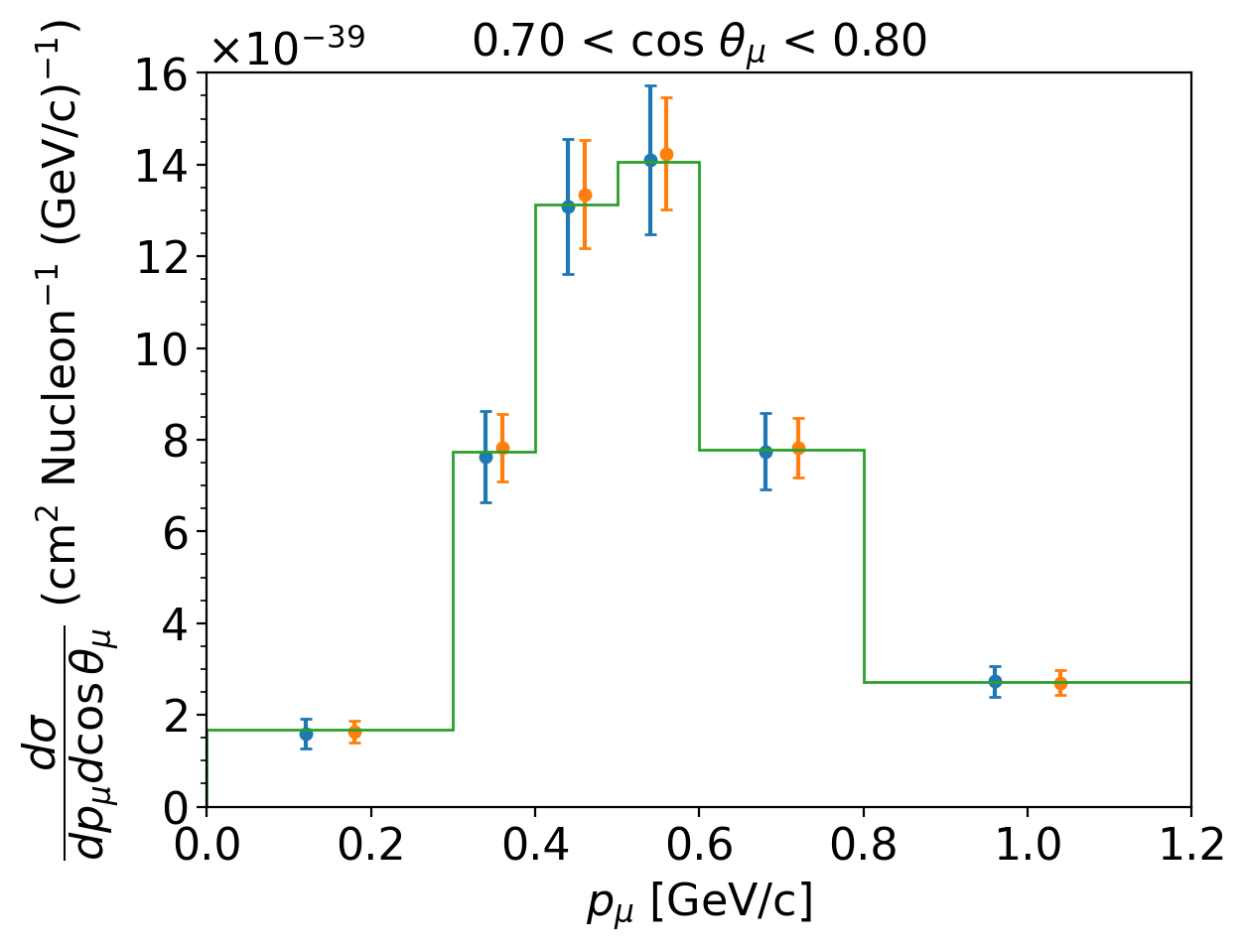}
\includegraphics[width=0.3\linewidth]{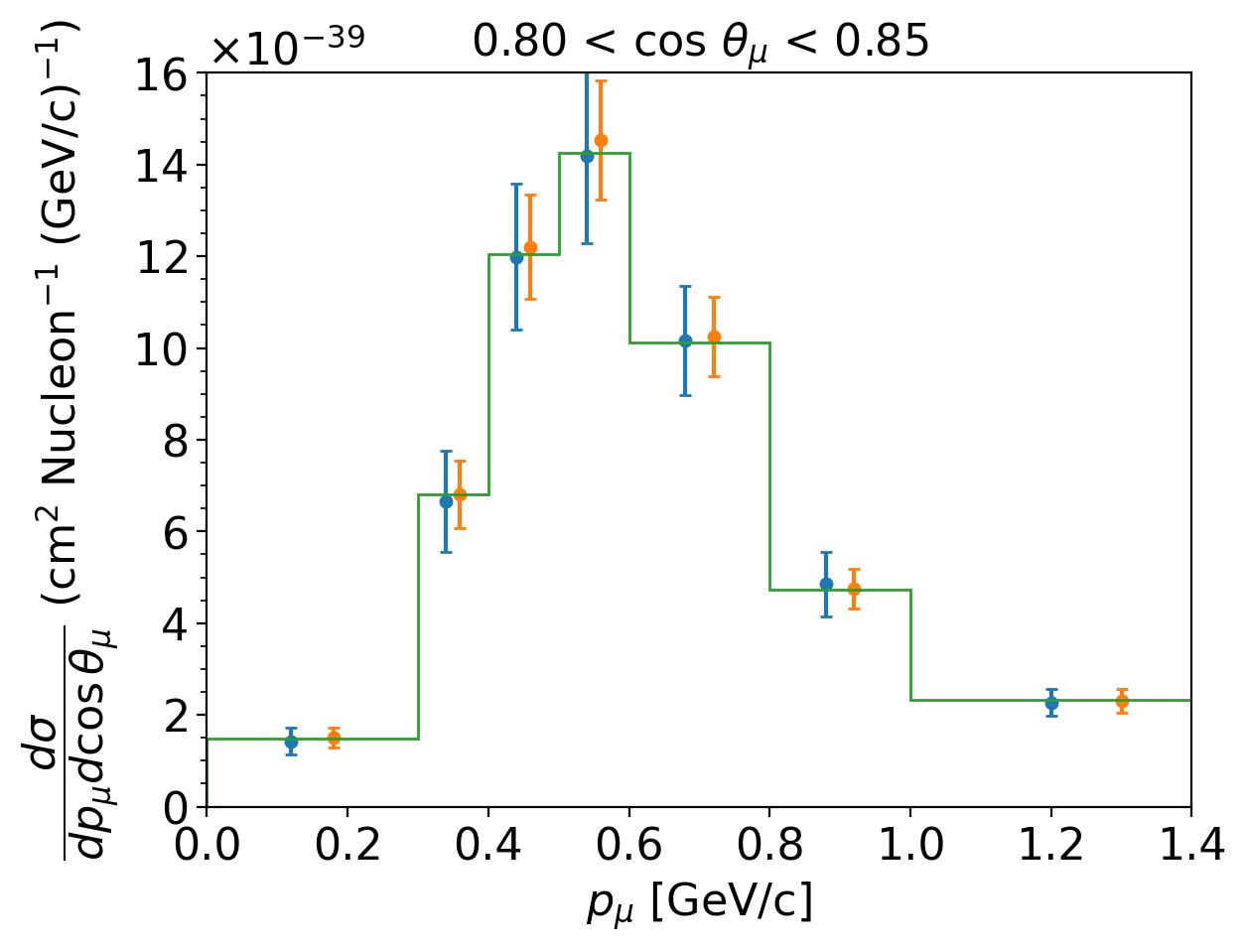}
\includegraphics[width=0.3\linewidth]{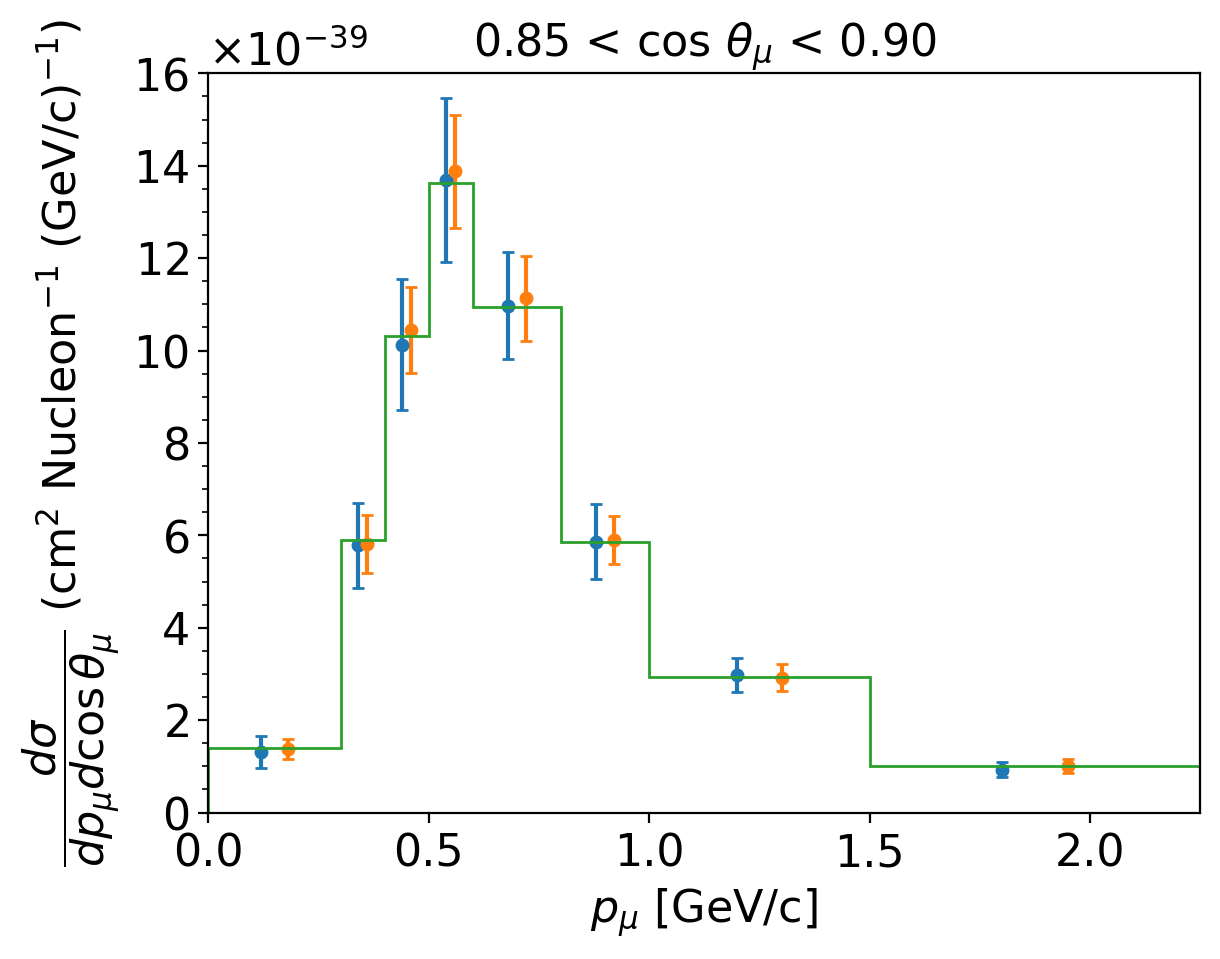}
\includegraphics[width=0.3\linewidth]{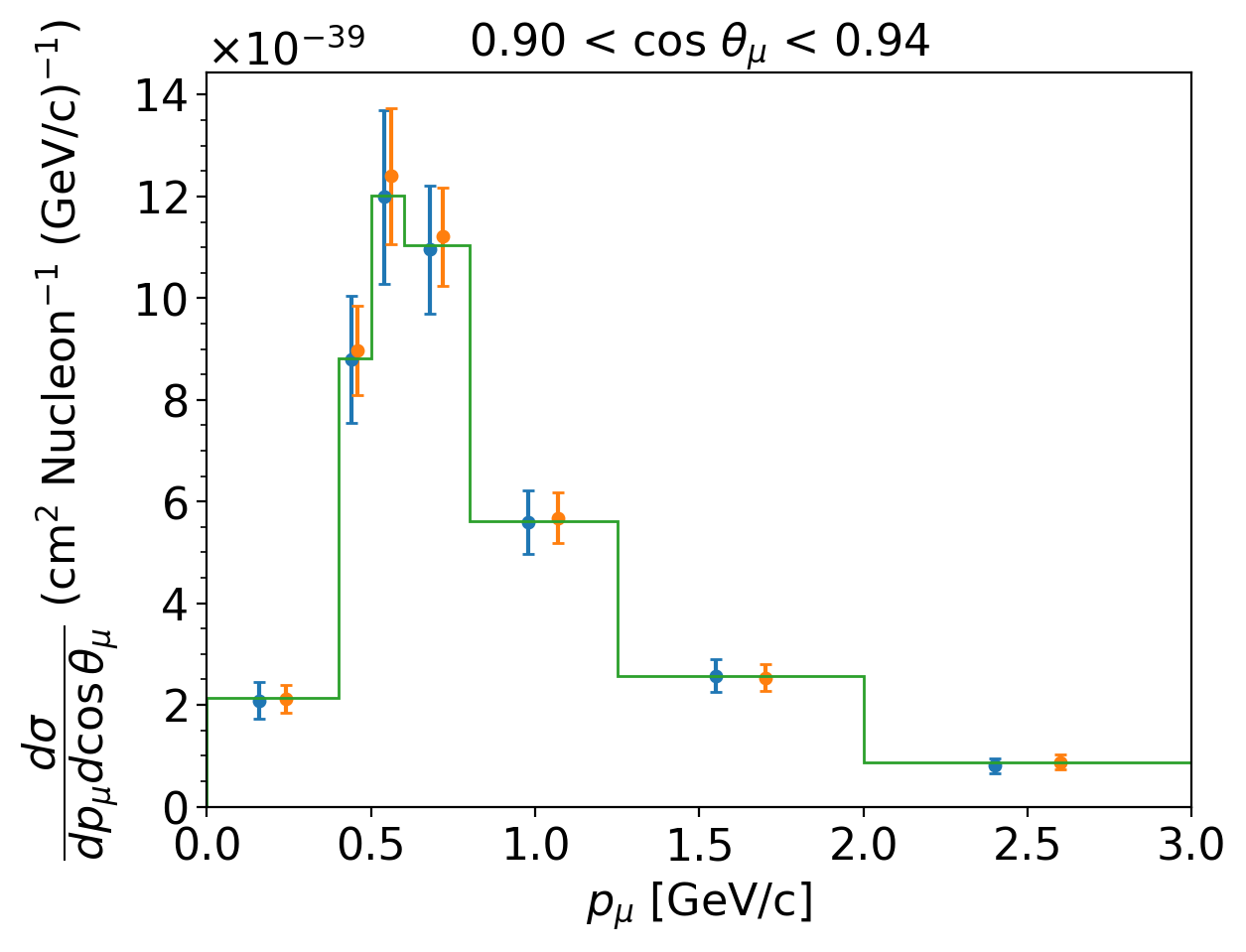}
\includegraphics[width=0.3\linewidth]{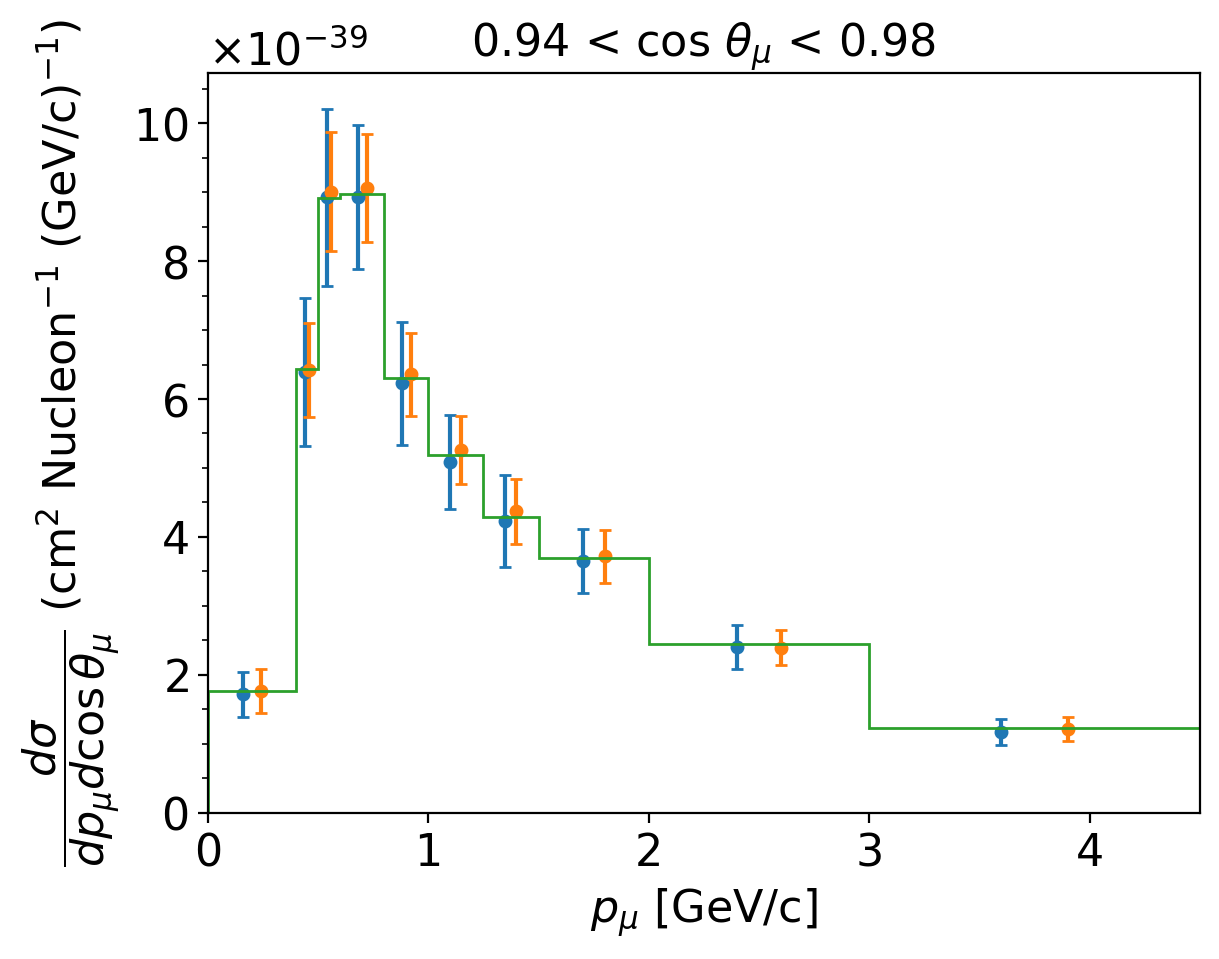}
\includegraphics[width=0.3\linewidth]{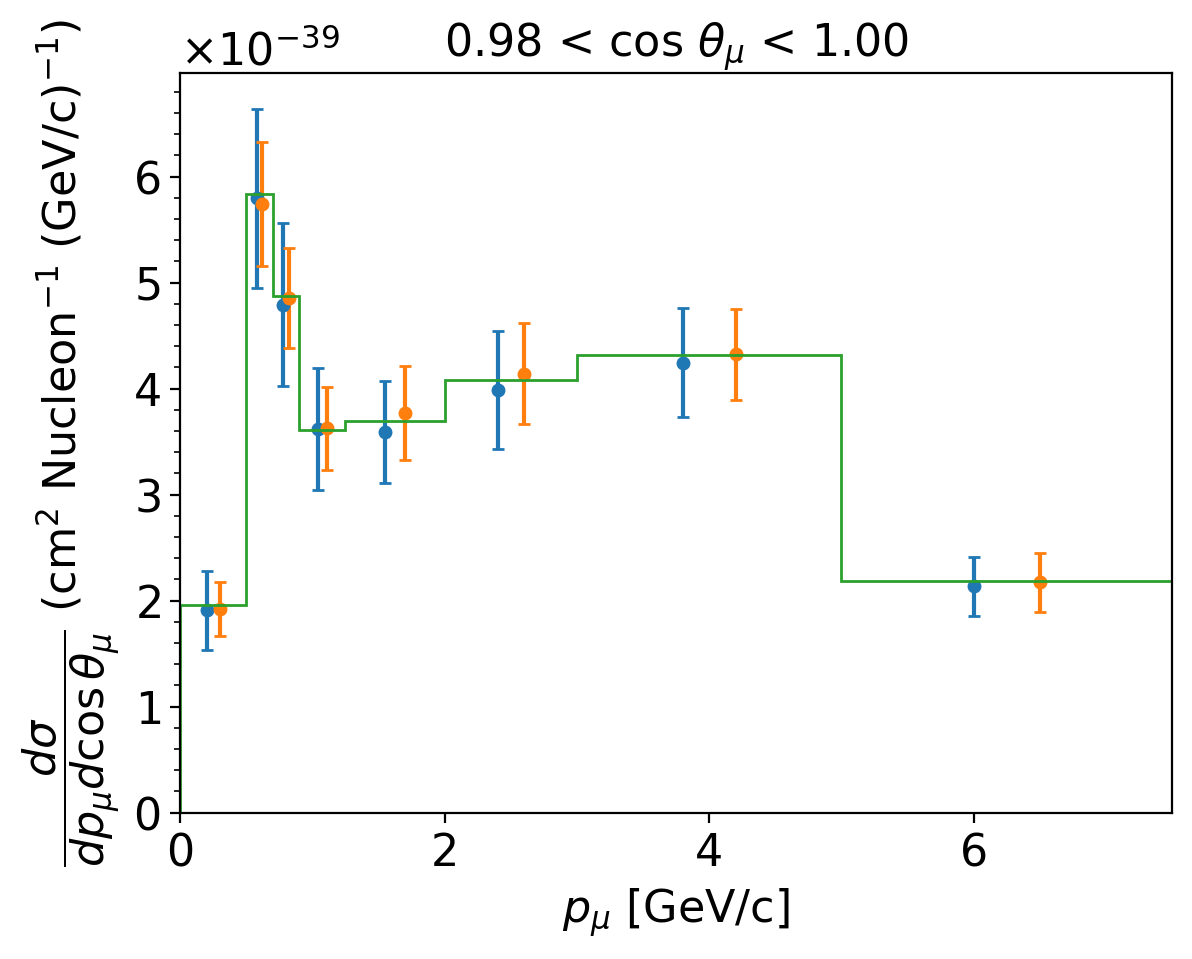}
\caption{\label{fig:UnfoldedMuonDistribution} Unfolded CC0$\pi$ differential cross sections as a function of the muon kinematics $(p_{\mu}, \cos \theta_{\mu})$, compared against the truth-level cross sections underlying the data. Each plot contains the differential cross sections from a specified range of the forward angle $\cos \theta_{\mu}$. Error bars are the spread in results from 500 pseudo-experiments varying systematic and statistical uncertainties, while the nominal result in each bin is the mean from those 500 unfolded throws. Note that the highest momentum bin in each case extends up to 30 GeV, but is truncated in the plots for readability.}
\end{figure}

\begin{figure}
\centering
\includegraphics[width=0.3\linewidth]{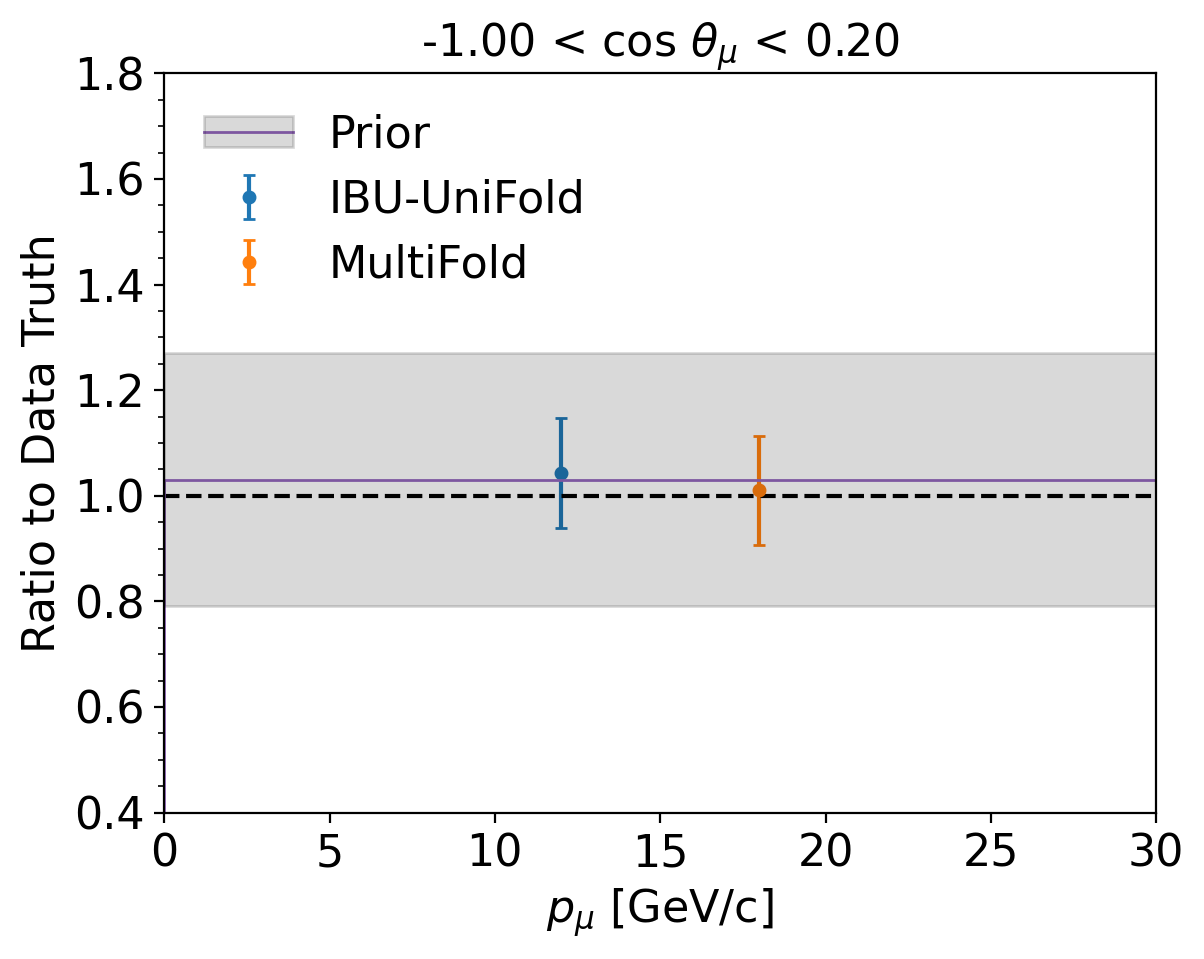}
\includegraphics[width=0.3\linewidth]{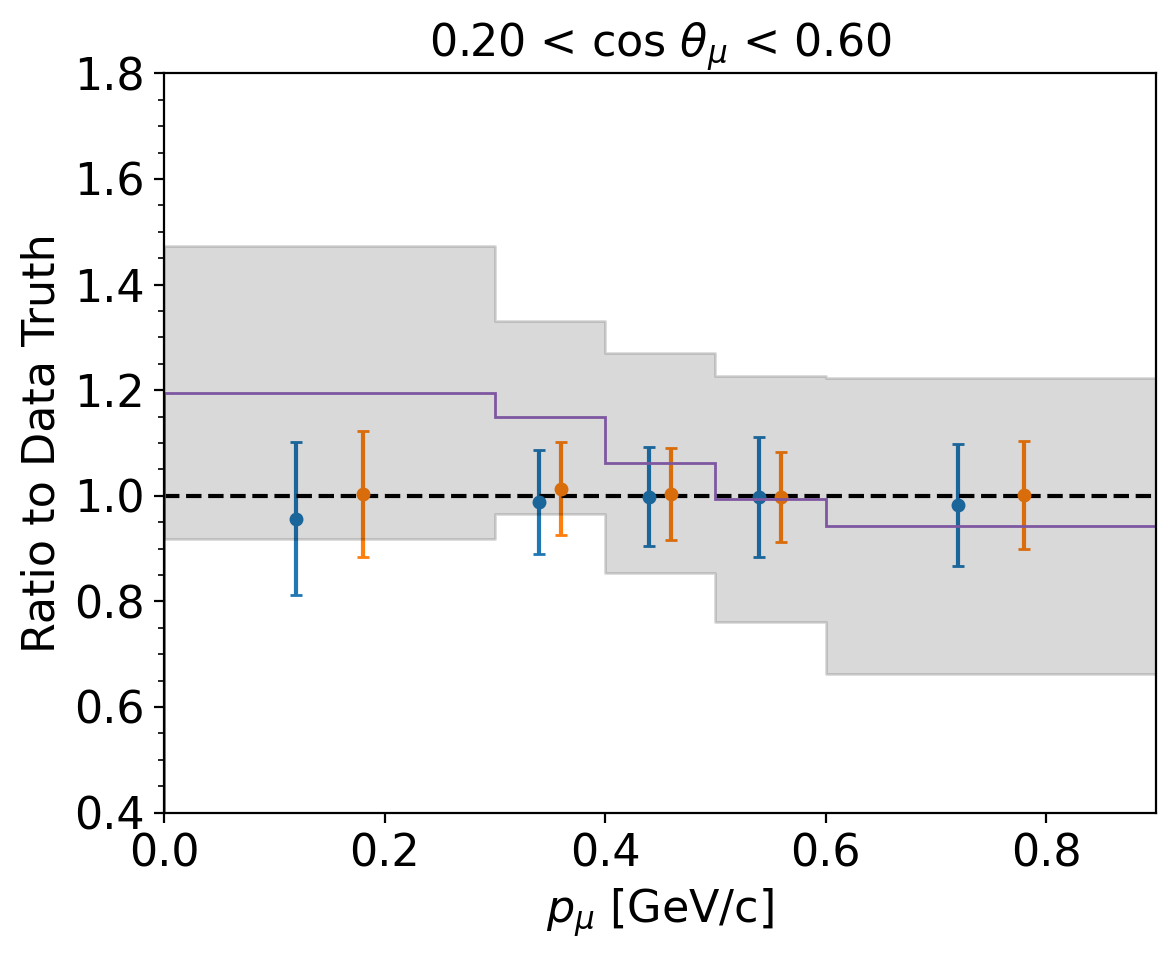}
\includegraphics[width=0.3\linewidth]{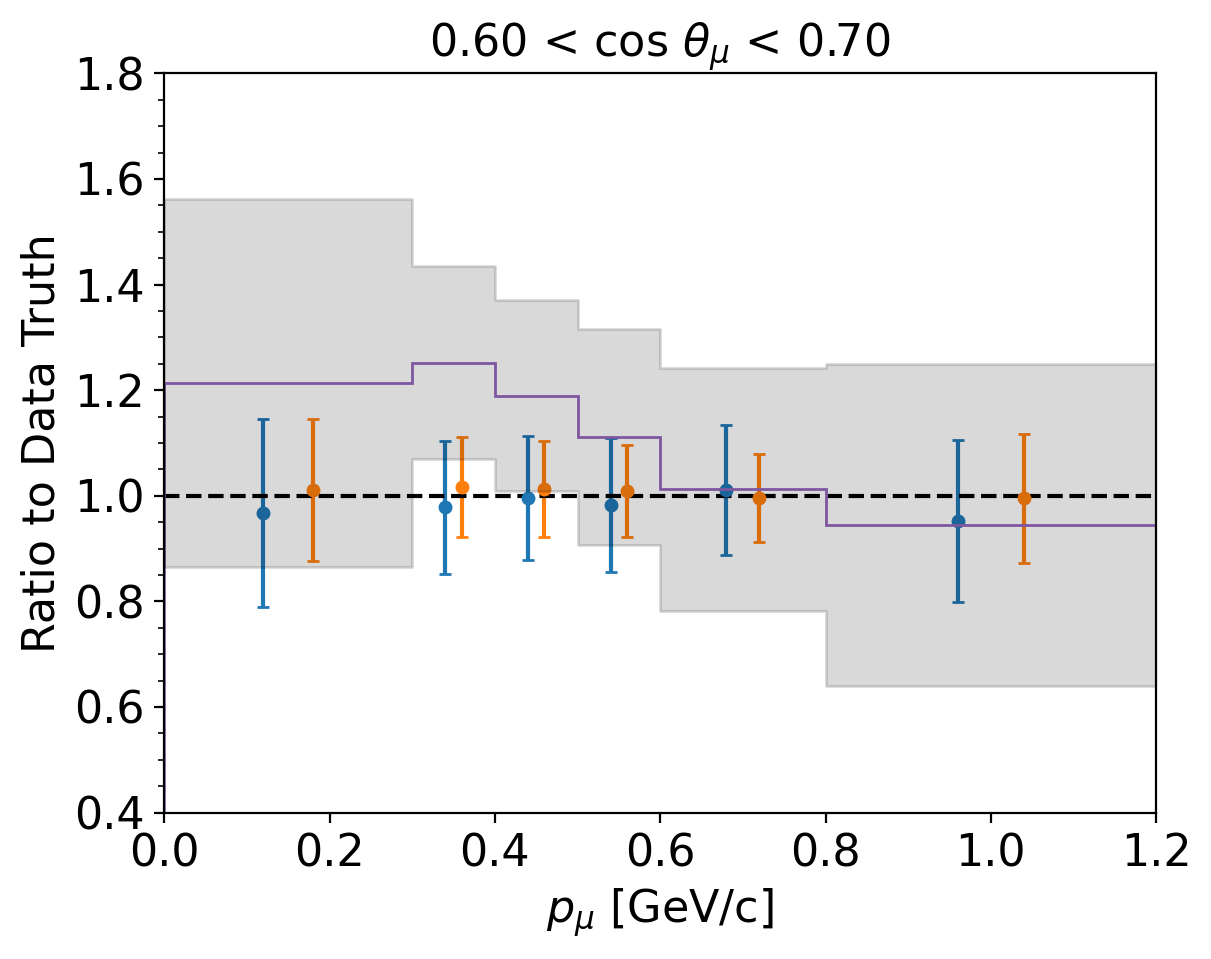}
\includegraphics[width=0.3\linewidth]{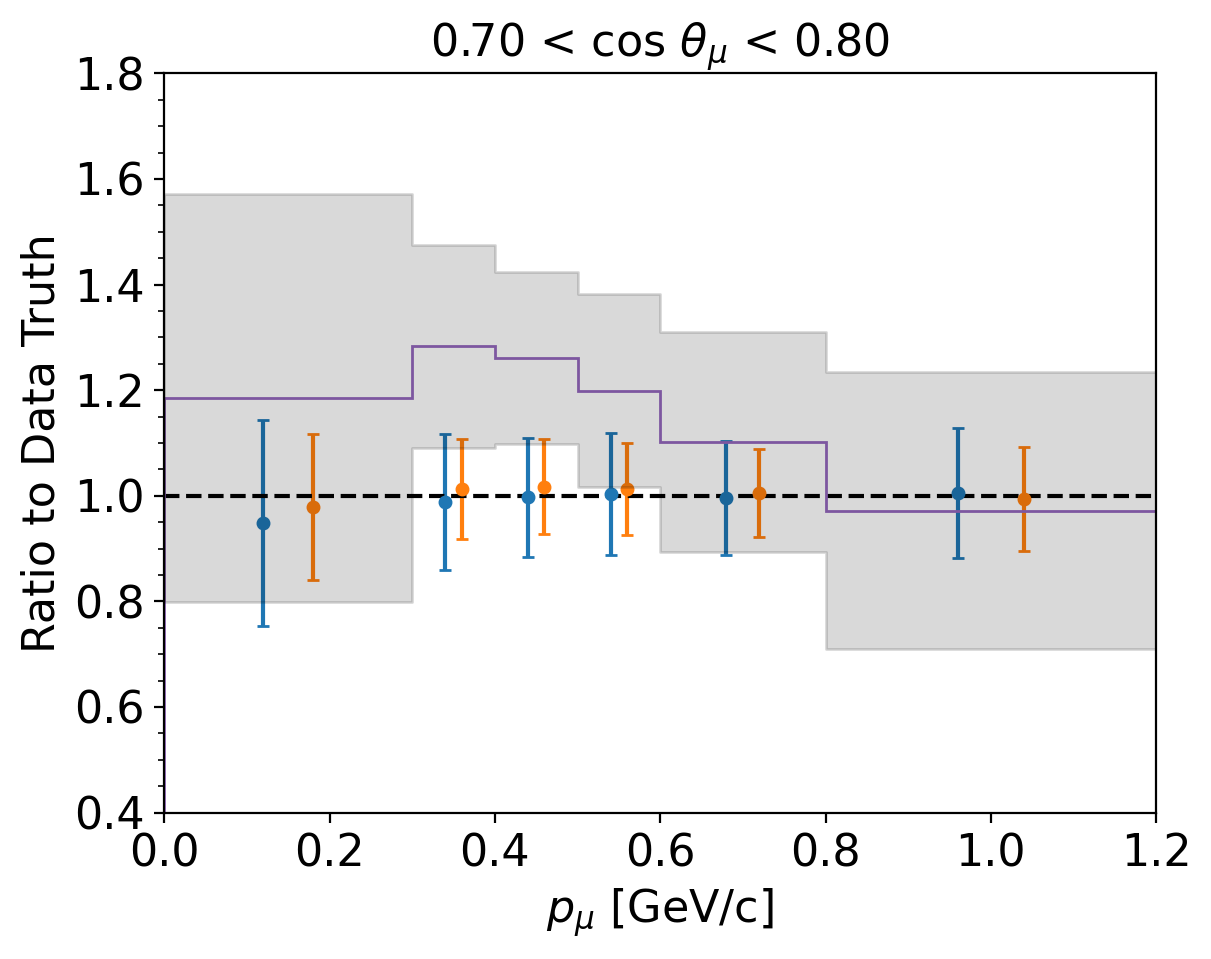}
\includegraphics[width=0.3\linewidth]{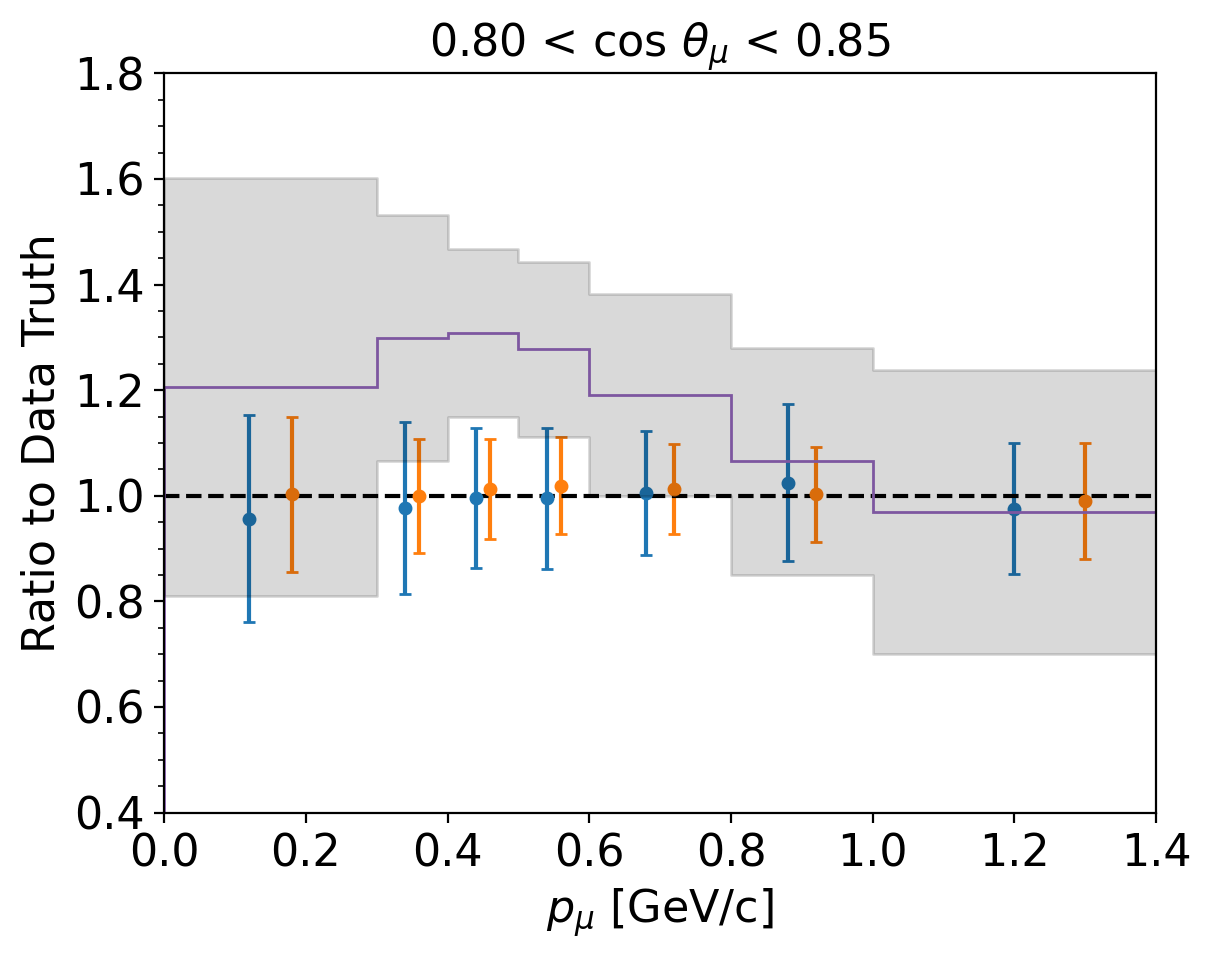}
\includegraphics[width=0.3\linewidth]{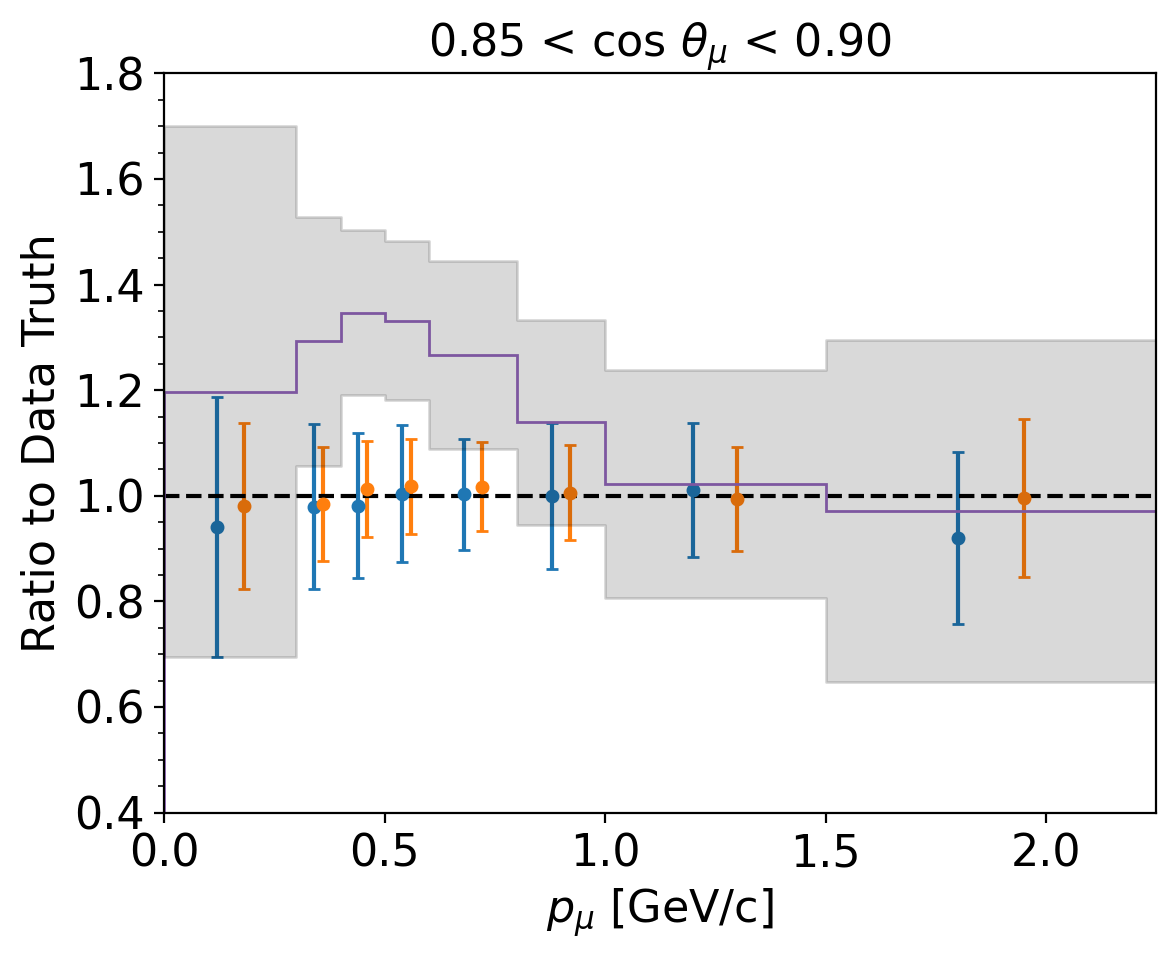}
\includegraphics[width=0.3\linewidth]{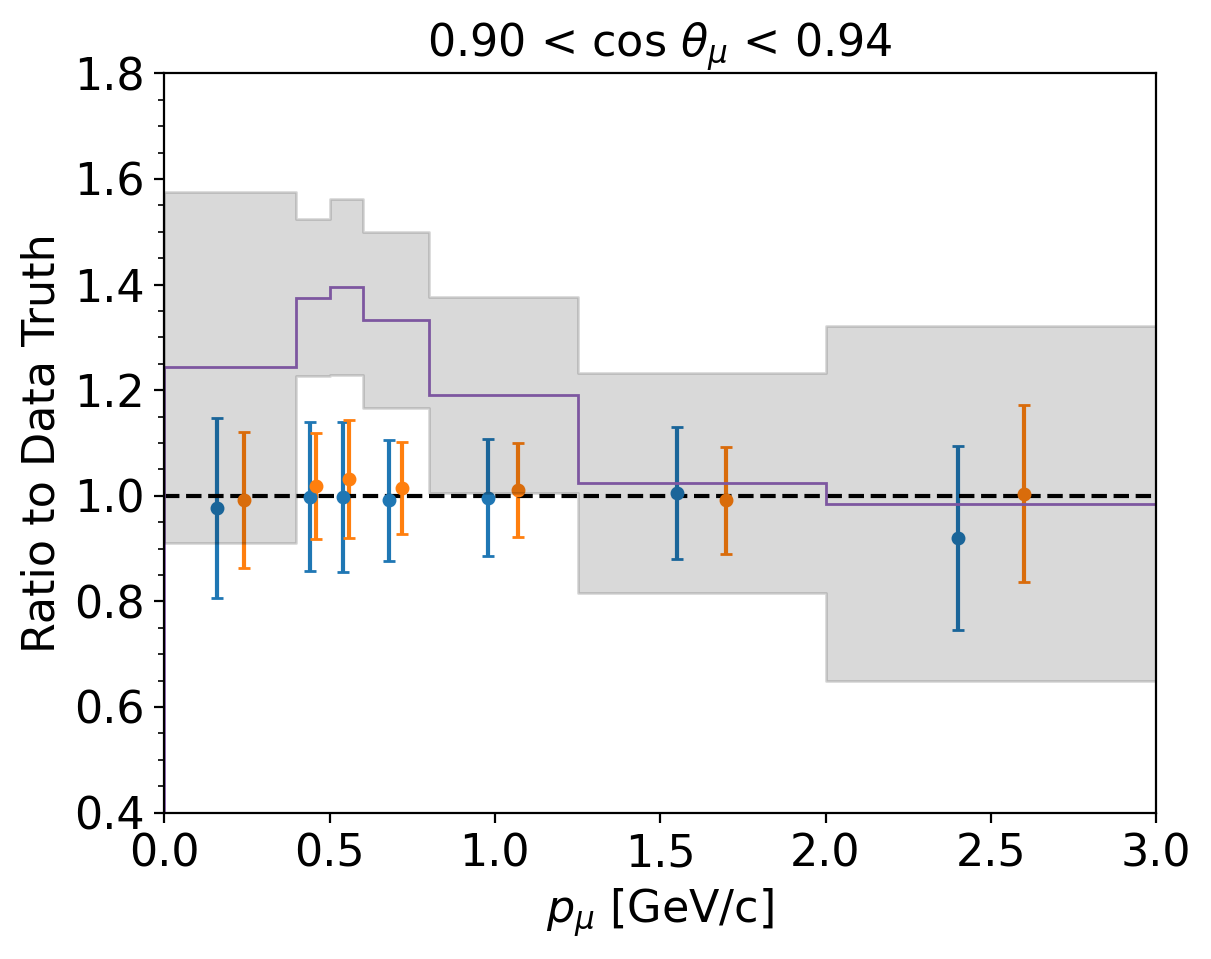}
\includegraphics[width=0.3\linewidth]{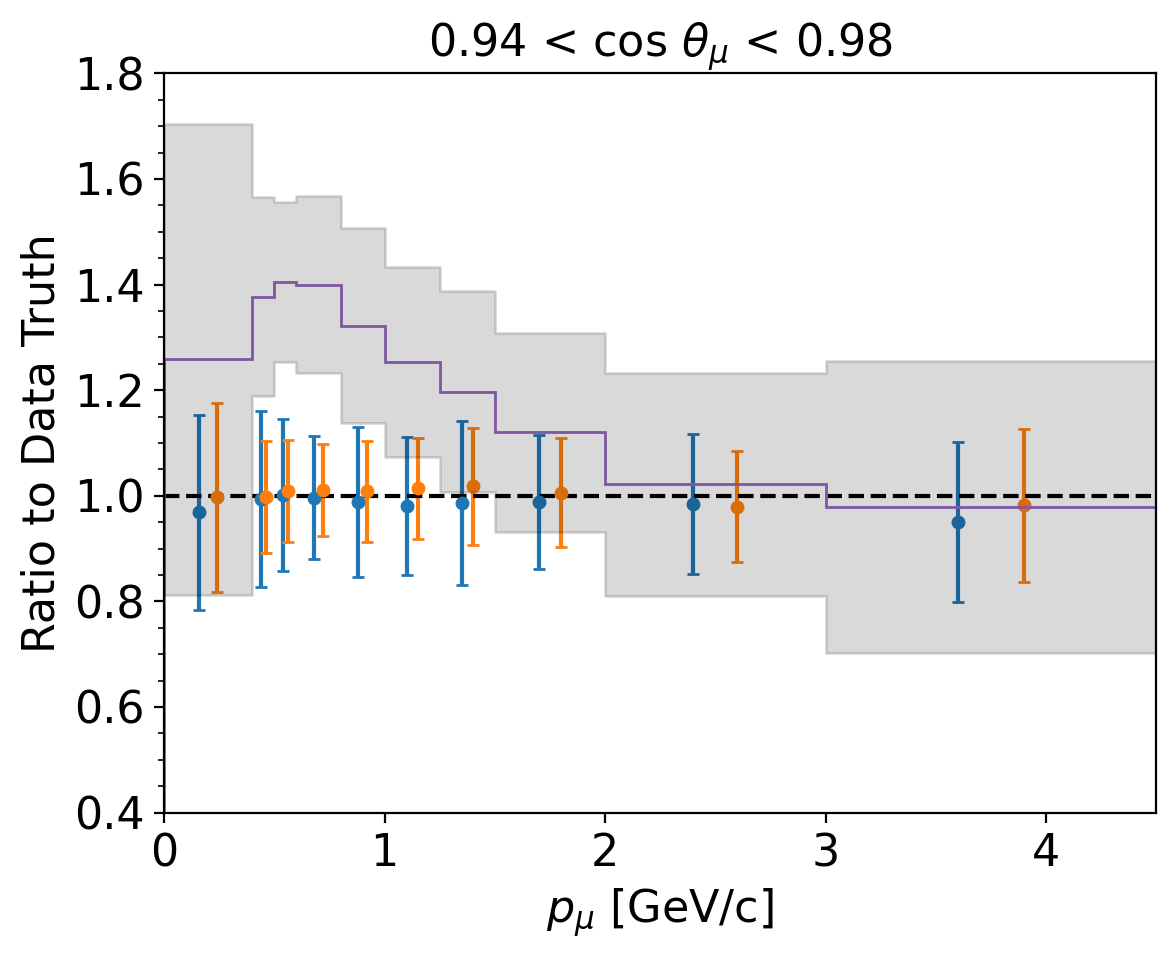}
\includegraphics[width=0.3\linewidth]{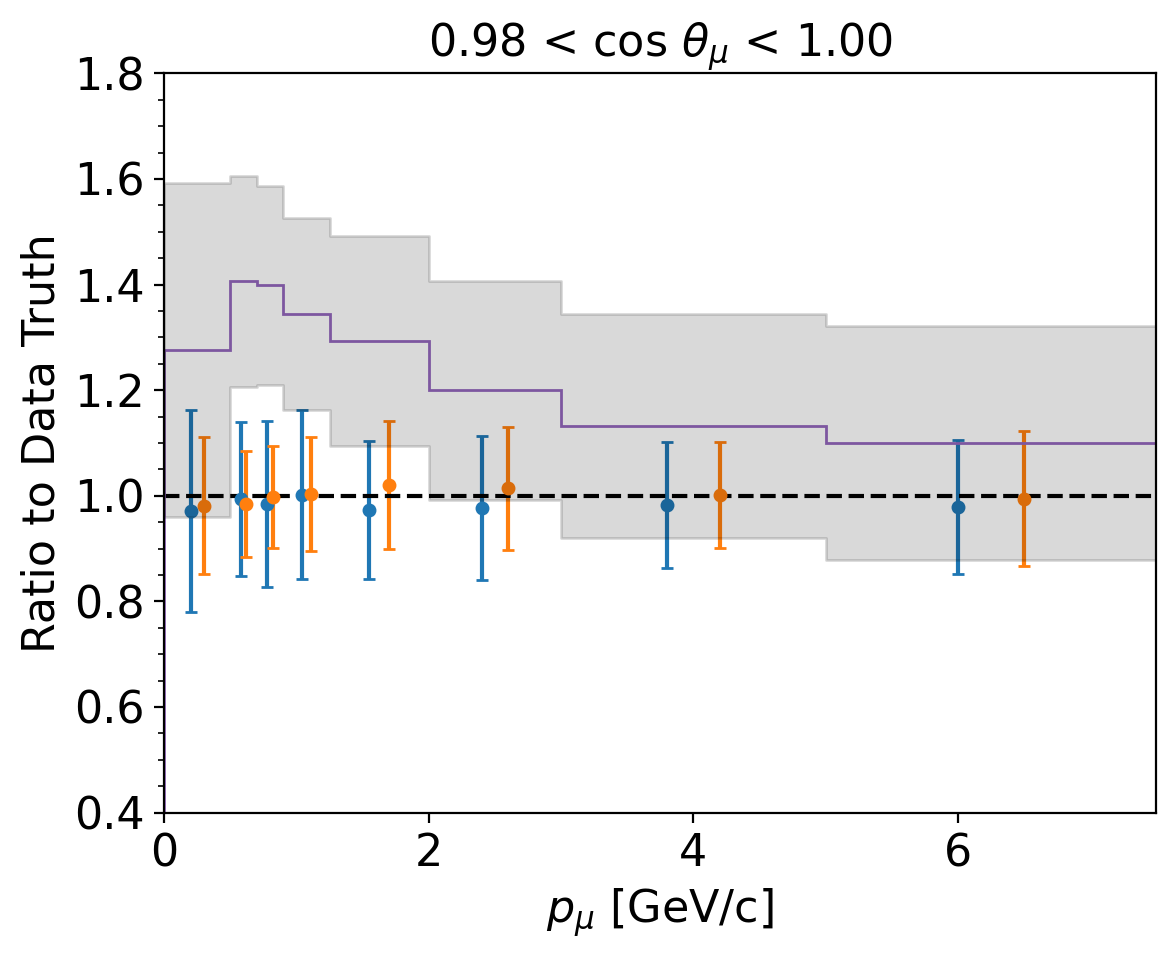}
\caption{\label{fig:UnfoldedMuonDistributionRatio} Ratios of the unfolded CC0$\pi$ differential cross sections as a function of the muon kinematics $(p_{\mu}, \cos \theta_{\mu})$, compared against the true cross sections used to generate the data. Each plot contains the ratios from a specified range of the forward angle $\cos \theta_{\mu}$. Error bars are the spread in results from 500 pseudo-experiments varying systematic and statistical uncertainties, while the nominal result for each bin is the mean of those 500 unfolded throws. Note that the highest momentum bin in each case extends up to 30 GeV, but is truncated in the plots for readability.}
\end{figure}

\begin{figure}
\centering
        \includegraphics[width=.42\linewidth]{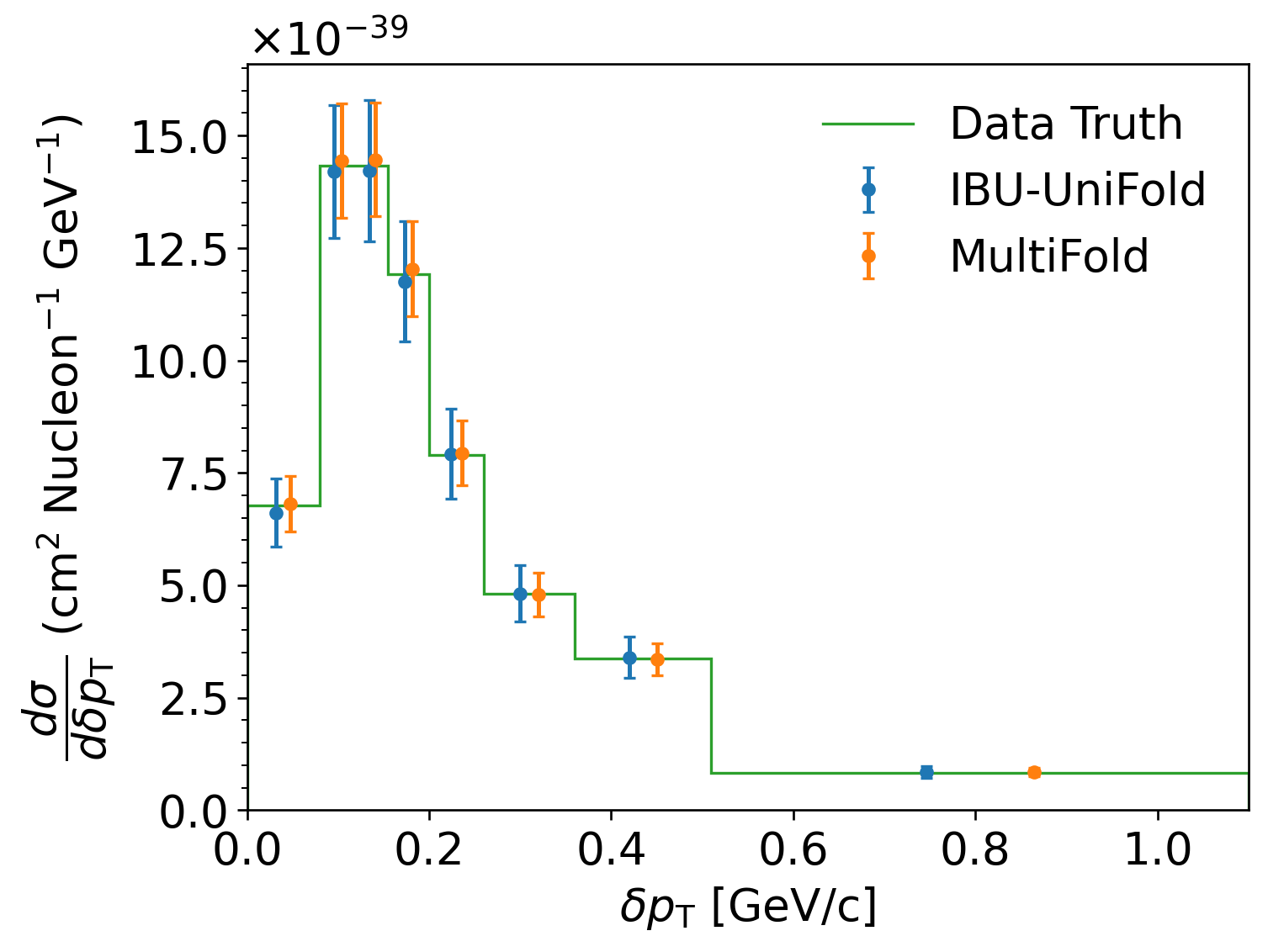}
        \includegraphics[width=.42\linewidth]{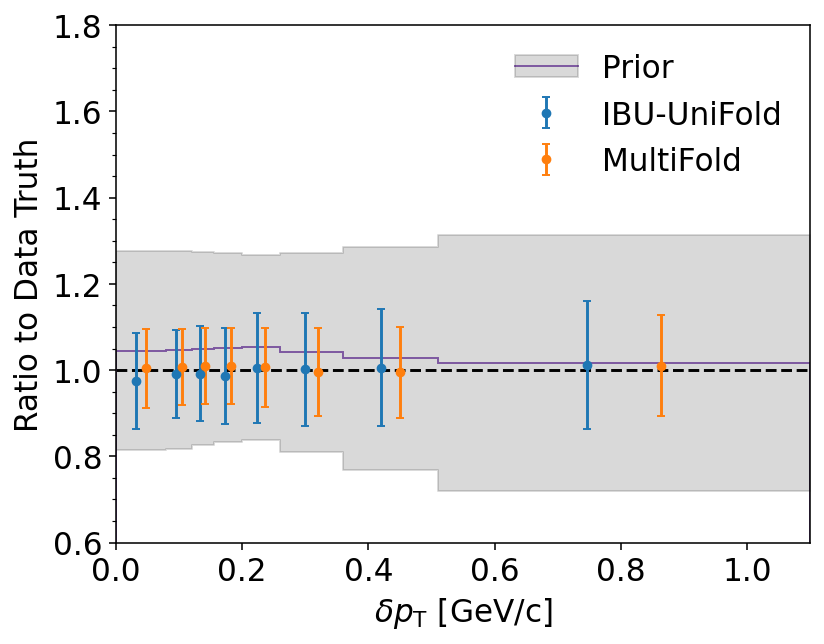}
        \includegraphics[width=.42\linewidth]{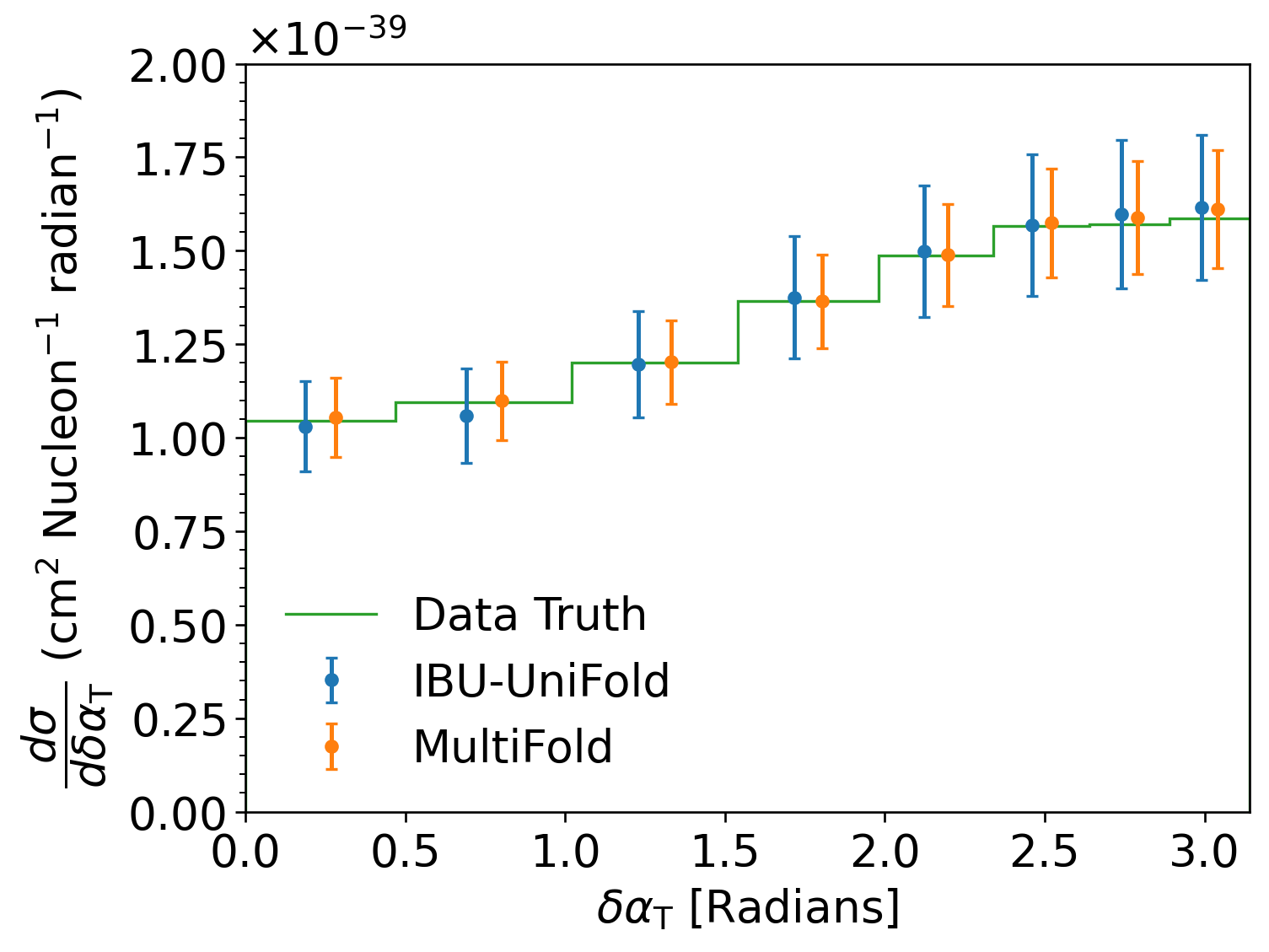}
        \includegraphics[width=.42\linewidth]{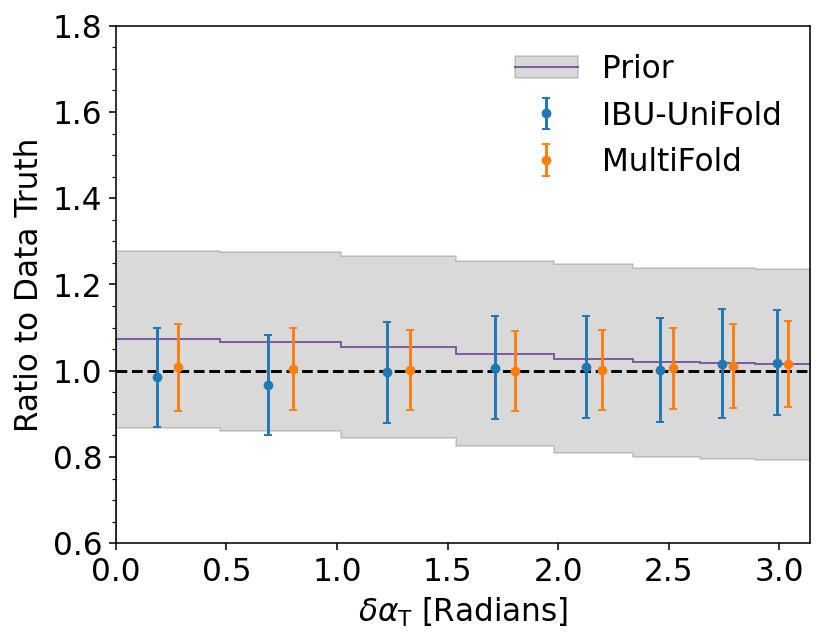}
        \includegraphics[width=.42\linewidth]{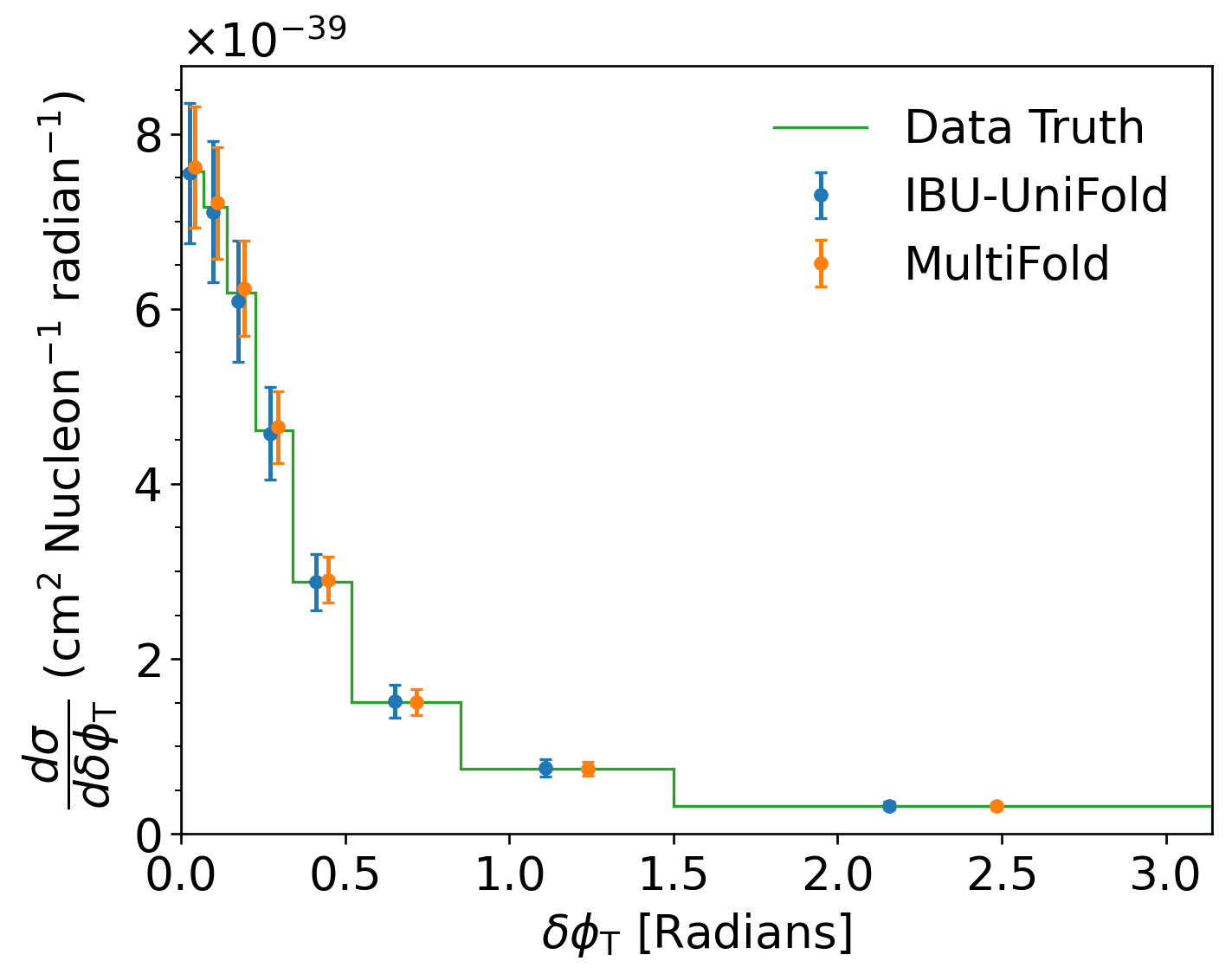}
        \includegraphics[width=.42\linewidth]{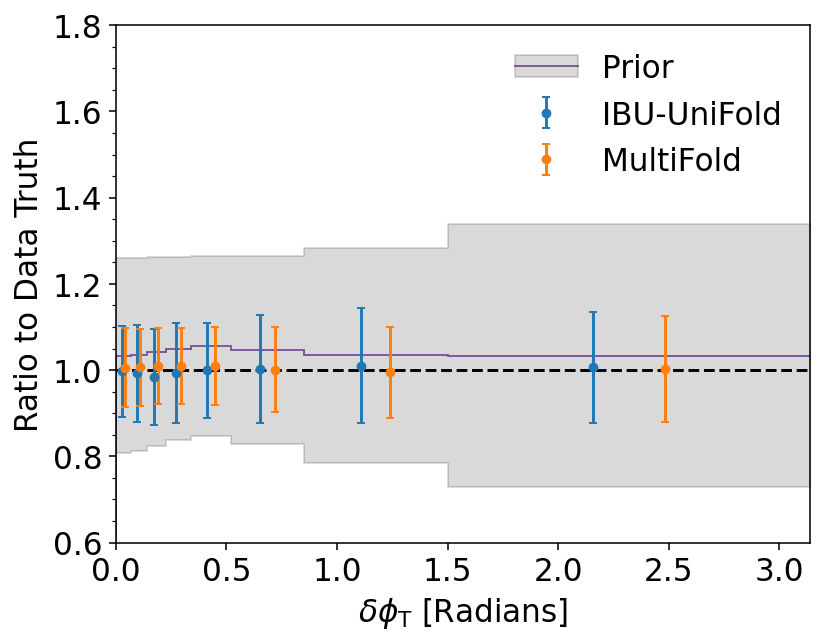}
    \caption{\label{fig:UnfoldedSTVDistributions} Unfolded CC0$\pi$ differential cross sections as functions of the three STVs, compared against the true cross sections underlying the data. Only events with a true proton momentum $> 450$ MeV are included for these comparisons. Error bars are the spread in results from 500 pseudo-experiments varying systematic and statistical uncertainties, and the nominal result in each bin is the mean from those 500 unfolded throws. \textbf{Left}: unfolded cross sections from IBU-UniFold and MultiFold compared against the true cross sections used to generate the data. \textbf{Right}: per-bin ratios of the unfolded cross sections from IBU-UniFold and MultiFold compared against the true cross sections for the data. The prior shows  the mean and standard deviation of each bin from the 500 pseudo-experiments before applying any unfolding procedure.}
\end{figure}

For direct visualization of the results, the unfolded differential cross sections from both IBU-UniFold and MultiFold compared against the true values underlying the data are shown for the muon kinematics in Fig. \ref{fig:UnfoldedMuonDistribution} and Fig. \ref{fig:UnfoldedMuonDistributionRatio}, and for the STVs in Fig. \ref{fig:UnfoldedSTVDistributions}.

\twocolumngrid

\section{Conclusions}
\label{sec:conclusions}

With its approach of event-by-event reweighting with the assistance of machine learning, OmniFold offers the ability to perform unbinned unfolding of multiple observables simultaneously while accounting for background and efficiency corrections with high-dimensional dependencies. We have demonstrated for the first time its application to a neutrino cross-section measurement, using public simulated data from the T2K ND280 detector. By creating fake data with nontrivial reweightings of the underlying interaction models, we were able to test the closure properties of the OmniFold approach, finding that with optimized inputs it was able to achieve similar $\chi^2$ goodness of fit with less bias and smaller uncertainties compared to a conventional IBU approach across several observables of interest.

Our results did not show strict improvement in terms of $\chi^2$ from using OmniFold-based approaches compared to a conventional unfolding method. Moving from a binned to unbinned approach without the addition of other information yielded generally worse results, which seems to stem from imperfect neural network training. We note that while we used a dense neural network for all setups in this study for better comparability, in practice a simpler classifier architecture such as a boosted decision tree would be more appropriate if one actually wanted to do a UniFold-like analysis with only one or two input quantities. We do see improvement as we move to higher-dimensional inputs, and comparing MultiFold against OmniFold, we see that directly providing the most relevant observables yields better performance than requiring the procedure to derive this information from other more general inputs, telling us that thoughtful choices of input variables are important. For any future application of OmniFold to real data, we thus emphasize the importance of using fake data studies to optimize the classifier architecture, training settings, and inputs. Some of the issues we observed are partially caused by the generally low statistics associated with a neutrino dataset, which will continue to be a problem for real neutrino data. Since OmniFold requires training with the actual data, improving the statistics is not a trivial matter of running more simulations, and this problem will require particular attention. Other methods for data augmentation in unfolding, including pre-training~\cite{Mikuni:2025tar,Mikuni:2024qsr}, would be useful to explore in future work.  

That being said, the performance of the OmniFold method was likely also limited by the nature of the public dataset that we used. For instance, the dataset only contained information about the primary muon and leading proton in each event, but we would expect OmniFold to do a better job of dealing with background contributions if we could provide pion kinematic information where it existed. For events where more than one proton is reconstructed, providing information about the subleading protons should be helpful too. This is supported by our general observation that our MultiFold and OmniFold implementations outperformed UniFold by making use of additional information besides just the unfolding variable. The public dataset that we used lacked the full stack of potentially useful information because it was not used in the conventional analysis, but for a future application of OmniFold to actual data we would want to maximize the amount of useful information that we provide as input. Our choice of a relatively simple neural network was also informed by the limited available information. More sophisticated classifier architectures could be used for datasets dealing with larger and variable-length particle stacks \cite{EnergyFlow}, although the available statistics will have to be considered when making this choice.

In conclusion, this work shows that OmniFold is a useful unfolding method to apply to neutrino measurements. Even with a limited dataset, it is able to deliver on its theoretical advantages, using high dimensional information to improve the quality of the unfolding and do this for multiple observables simultaneously. While we used a T2K ND280 dataset, the advantages should be similar for any detector. We have identified the difficulties encountered during this work, some of which may be particular to neutrino datasets and will require further study, but OmniFold shows promise as an unfolding technique for all neutrino experiments to improve the quality of their measurements. 

\begin{acknowledgements}
We would like to thank T2K for making the simulated dataset extensively used in this work publicly available. We would also like to particularly thank Ciro Riccio for his invaluable help in understanding the T2K public dataset and in his efforts to produce the dataset initially. Additionally, we would like to thank Ryan Milton for validating the implementation of OmniFold used in this work with his RooUnfold implementation.
The work of RH, BN and CW was supported by the U.S. Department of Energy, Office of Science, Office of High Energy Physics, under contract number DE-AC02-05CH11231.
The work of MK and TK was supported by JSPS KAKENHI Grant Number JP23H04504.
The work of AC was supported by the U.S. Department of Energy, Office of Science, Office of High Energy Physics, under DOE award DE-SC0010005. 
Additional support was provided through the U.S.‐Japan Science and Technology Cooperative Research Program in High Energy Physics.
This research used resources of the National Energy Research Scientific Computing Center (NERSC), a U.S. Department of Energy Office of Science User Facility located at Lawrence Berkeley National Laboratory, operated under Contract No. DE-AC02-05CH11231 using NERSC awards ERCAP0023703, ERCAP0028625 and ERCAP0031852.
\end{acknowledgements}

\FloatBarrier
\bibliographystyle{apsrev4-1}

\onecolumngrid

\appendix

\section{Correlation Matrices}

\begin{figure}
    \includegraphics[width=0.45\linewidth]{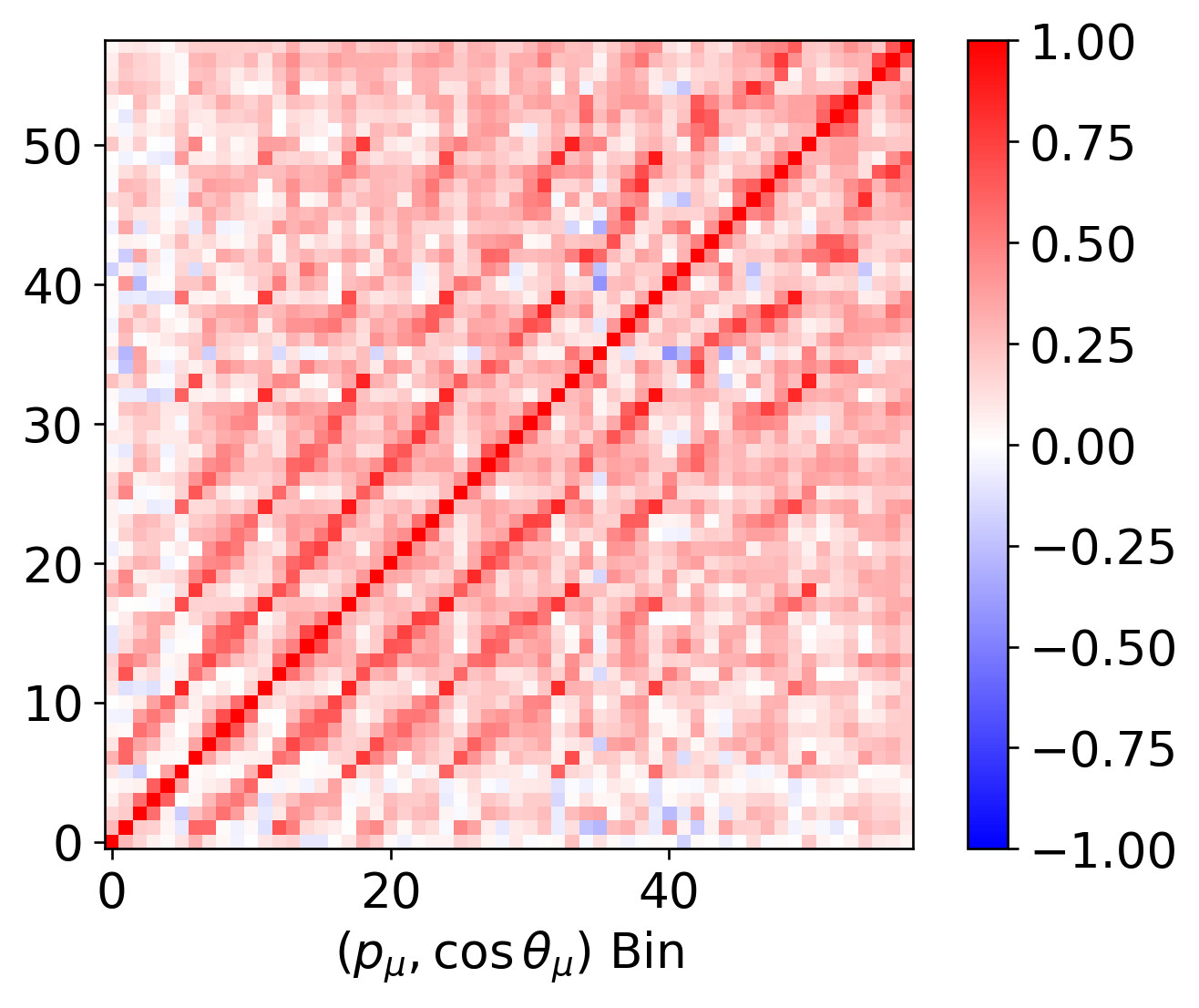}
    \includegraphics[width=0.45\linewidth]{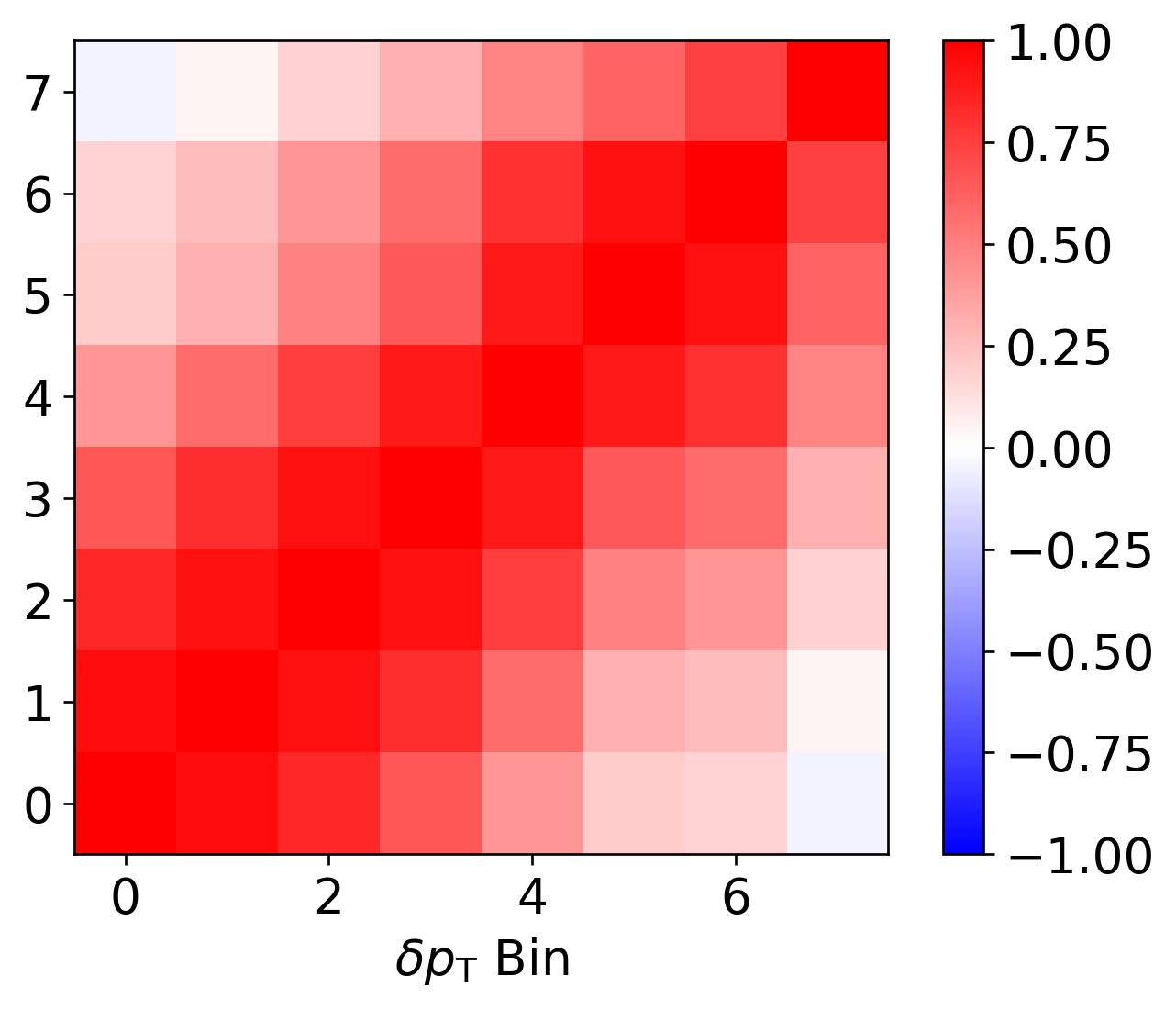}
    \includegraphics[width=0.45\linewidth]{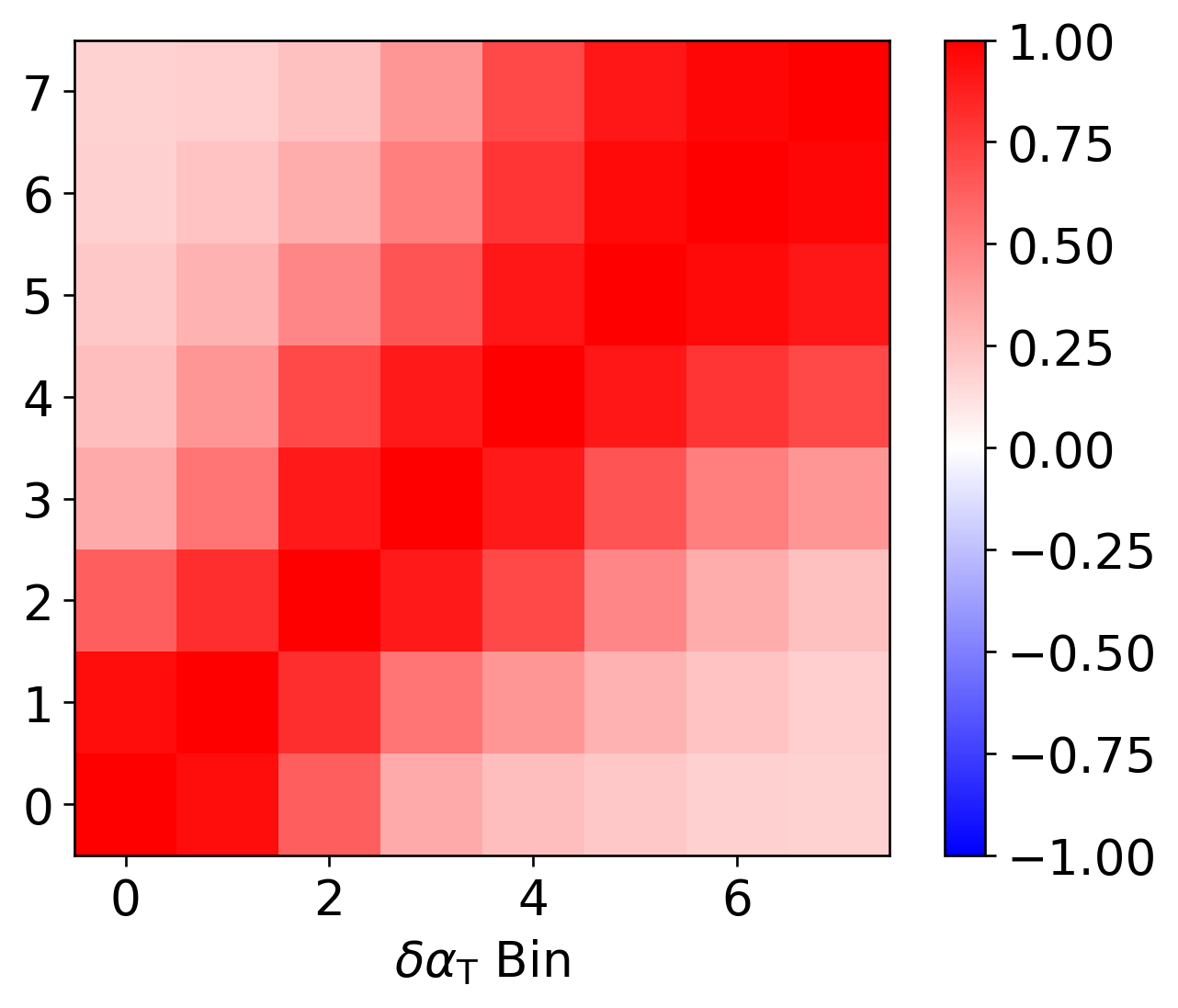}
    \includegraphics[width=0.45\linewidth]{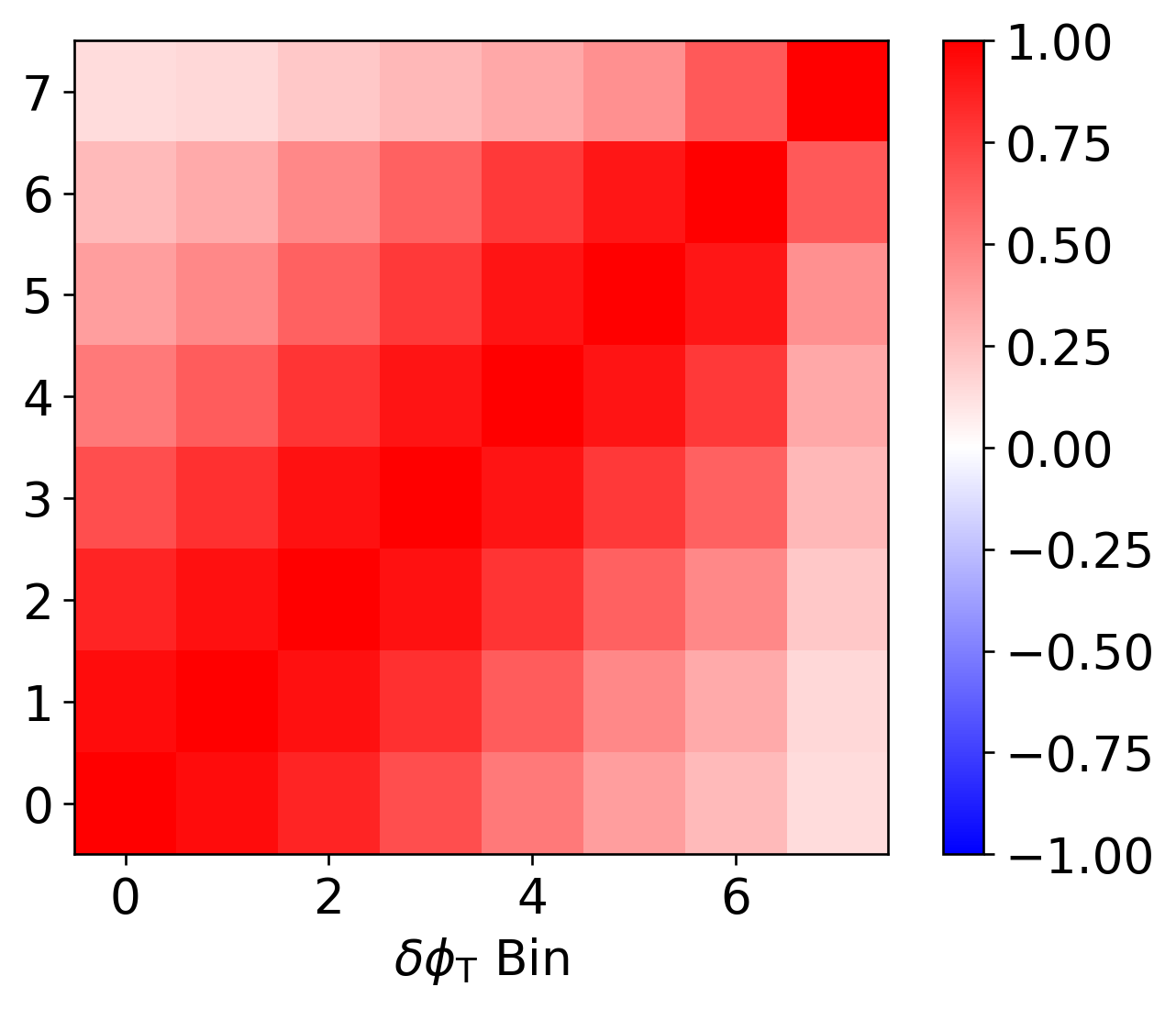}
    \caption{\label{fig:CorrelationMatrices} Correlation matrices from the MultiFold method for the number of events per bin in the unfolded distributions of each observable of interest. The correlations in the results are roughly as expected given the nature of the systematic uncertainties applied to the simulation.}
\end{figure}

Correlation matrices for the MultiFold result are provided in Fig. \ref{fig:CorrelationMatrices}, for each of the observables of interest. The mapping of the bin indices to actual bin values can be found in Tab. \ref{tab:MuonBinning} and Tab. \ref{tab:STVBinning}. These correlation matrices are calculated using the unfolded results from 500 throws of the simulation that account for systematic and statistical uncertainties, but without accounting for flux normalization. Flux uncertainties would otherwise add a notable positive correlation to all bins in the result.

\onecolumngrid

\end{document}